\documentclass[11pt]{article}
\usepackage{a4,bm,epsfig}
\usepackage{ifthen,array}
\newcommand{\ams}{\usepackage{amsfonts,amssymb,amsmath}}

\ams
\allowdisplaybreaks[3]
\newlength{\textwidthorig}
\newlength{\oddsidemarginorig}
\newlength{\textheightorig}
\newlength{\topmarginorig}
\def\seitenlaengenabsolut#1 #2 #3 #4 {\setlength{\textwidth}{#1}
                                      \setlength{\oddsidemargin}{#2}
                                      \setlength{\textheight}{#3}
                                      \setlength{\topmargin}{#4}}
\def\seitenlaengenrelzustandard#1 #2 #3 #4 {\setlength{\textwidth}{\textwidthorig+#1}
                                            \setlength{\oddsidemargin}{\oddsidemarginorig+#2}
                                            \setlength{\textheight}{\textheightorig+#3}
                                            \setlength{\topmargin}{\topmarginorig+#4}}
\def\seitenlaengenrelzuvorher#1 #2 #3 #4 {\addtolength{\textwidth}{#1}
                                          \addtolength{\oddsidemargin}{#2}
                                          \addtolength{\textheight}{#3}
                                          \addtolength{\topmargin}{#4}}
\newcommand{\standardseite}{\seitenlaengenrelzuvorher2.2cm -0.8cm 1.8cm -1.5cm }   %
\standardseite

\newlength{\laengespatium}

\newcommand{\einbett}{\hookrightarrow}   
\newcommand{\nach}{\longrightarrow}      

\newcommand{\auf}{\longmapsto}           
\newcommand{\txtauf}[1]{\auf}            
\newcommand{\impliz}{\Longrightarrow}    
\newcommand{\aequ}{\Longleftrightarrow}  
 
\newcommand{\invimpliz}{\Longleftarrow}  
\newcommand{\gegen}{\rightarrow}         
\newcommand{\iso}{\cong}                 
\newcommand{\kong}{\ident}               
\newcommand{\ident}{\equiv}              
\newcommand{\teilmenge}{\subseteq}       
\newcommand{\obermenge}{\supseteq}       
\newcommand{\echteteilmenge}{\subset}    
\newcommand{\aeqrel}{\sim}               
\newcommand{\nichtin}{\not\in}

\newcommand{\esex}{\exists}              
\newcommand{\leeremenge}{\varnothing}    
\newcommand{\tensor}{\otimes}            
\newcommand{\kreuz}{\times}              
\newcommand{\grosskreuz}{\mbox{\Large $\boldsymbol{\kreuz}$}}
\newcommand{\einschr}[1]{{}\arrowvert_{#1}}      

\newcommand{\betraganpass}[1]%
           {\left| #1 \right|}           
\newcommand{\bigbetrag}[1]%
           {\bigl|{#1}\bigr|}            
\newcommand{\betrag}[1]%
           {|{#1}|}                      
\newcommand{\betragnichtanpass}[1]%
           {\mid #1 \mid}                
\newcommand{\norm}[1]%
           {{}{\parallel}#1{\parallel}{}}      
\newcommand{\erww}[1]%
           {\langle #1 \rangle}          
\newcommand{\skalprod}[2]%
           {\langle #1,#2 \rangle}       
\newcommand{\quer}{\overline}            
\newcommand{\dach}{\widehat}             
\newcommand{\im}{\text{im\;}}                          
\newcommand{\inter}{\text{int}\:}                      
\newcommand{\spann}{\text{span}}                       
\newcommand{\supp}{\text{supp }}                       
\newcommand{\elanz}{\#}                                
\newcommand{\Hom}{\text{Hom}}                          
\newcommand{\dd}{\text{d}}                             
\newcommand{\field}[1]{\mathbb{#1}}                    
\newcommand{\K}{{\field{K}}}                           
\newcommand{\C}{{\field{C}}}                           
\newcommand{\N}{{\field{N}}}                           
\newcommand{\R}{{\field{R}}}                           
\newcommand{\Z}{{\field{Z}}}                           
\newcommand{\rnkl}[2]{\raisebox{-0.4ex}{$#1$}%
\raisebox{-0.12ex}{{\large$\setminus$}}\,#2}   
\newcommand{\agb}{{\overline{{\cal A}/{\cal G}}}}      
\newcommand{\agbfact}[1][]{\text{$\agb/\!\aeqrel$}}    
\newcommand{\Ab}{{\overline{{\cal A}}}}                
\newcommand{\A}{{\cal A}}                              
\newcommand{\Gb}{{\overline{{\cal G}}}}                

\newcommand{\qa}{{\quer{A}}}                           

\newcommand{\holgr}{{\mathbf H}}                       
\newcommand{\bz}{{\mathbf B}}                          

\newcommand{\gross}[1]{{\boldsymbol #1}}               
\newcommand{\gc}{\gross{\gamma}}                       
\newcommand{\gd}{\gross{\delta}}                       
\newcommand{\Pf}{{\cal P}}                             
\newcommand{\LG}{{\mathbf{G}}}                         
\newcommand{\LN}{{\mathbf{N}}}                         
\newcommand{\Lieg}{{\mathfrak{g}}}                            
\newcommand{\aeqrelzush}[1][]{\sim}                    

\newcommand{\nklza}[1][]{\ifthenelse{\equal{#1}{}}     
                                    {\rnkl{Z(\holgr_\qa)}{\LG}}        
                                   {\rnkl{Z(\holgr_{#1})}{\LG}}}       
\newcommand{\nkla}[1][]{\ifthenelse{\equal{#1}{}}      
                                    {\rnkl{\bz(\qa)}{\Gb}}        
                                    {\rnkl{\bz(#1)}{\Gb}}}       
\newcommand{\he}{{\text{he}}}                          
\newcommand{\ab}{{\text{ab}}}                          

\newcommand{\YM}{{\text{YM}}}                          

\newcommand{\ymwirk}[1][]{\ifthenelse{\equal{#1}{}}{S_{\YM}}{S_{\YM,#1}}}

\newcommand{\bmat}{\begin{pmatrix}}
\newcommand{\emat}{\end{pmatrix}}

\newcommand{\codim}{\text{codim}}
\newcommand{\ListNullAbstaende}{\setlength{\topsep}{0pt}%
                                \setlength{\parskip}{0pt}%
                                \setlength{\partopsep}{0pt}%
                                \setlength{\itemsep}{0pt}%
                                \setlength{\parsep}{0pt}}
\newcommand{\ListNurAnstrichAbstand}{\setlength{\topsep}{0pt}%
                                     \setlength{\parskip}{0pt}%
                                     \setlength{\partopsep}{0pt}%
                                     \setlength{\parsep}{0pt}}
\newenvironment{StandardListe}[2]%
               {\begin{list}%
                      {#1}%
                      {\settowidth{\leftmargin}{M#1}%
                       \settowidth{\labelwidth}{#1}%
                       \settowidth{\labelsep}{M}%
                       #2%
                      }%
                }%
               {\end{list}}%
\newenvironment{EinfachListe}[1]%
               {\begin{StandardListe}{#1}{\ListNullAbstaende}}%
               {\end{StandardListe}}%
               {\begin{StandardListe}{#1}{\ListNurAnstrichAbstand}}%
               {\end{StandardListe}}%
\newcommand{\labelsatz}[1]{#1}
\newcounter{listennr}                      %
\newlength{\hilfslaenge}
\newlength{\stdlabellaenge}
\newlength{\maximum}
\newcommand{\stdlabel}{}
\newcommand{\Maximum}{}
\newcommand{\iitem}[1][]{\ifthenelse{\equal{#1}{}}%
                           {\item \setlength{\hilfslaenge}{\stdlabellaenge}}%
                           {\item[\labelsatz{#1}\hfill]%
                            \settowidth{\hilfslaenge}{\labelsatz{#1}}}%
                         \ifthenelse{\lengthtest{\maximum < \hilfslaenge}}%
                           {\setlength{\maximum}{\hilfslaenge}%
                            \ifthenelse{\equal{#1}{}}%
                               {\renewcommand{\Maximum}{\stdlabel}}%
                               {\renewcommand{\Maximum}{#1}}}%
                           {}%
                      }      
\makeatletter
\newenvironment{AutoLabelLaengenListe}[2][]%
               {\begin{list}%
                      {\labelsatz{#1}\hfill}%
                      {\stepcounter{listennr}%
                       \settowidth{\leftmargin}{M\labelsatz{\ref{listnr\arabic{listennr}}}}%
                       \settowidth{\labelwidth}{\labelsatz{\ref{listnr\arabic{listennr}}}}%
                       \settowidth{\labelsep}{M}%
                       \settowidth{\stdlabellaenge}{\labelsatz{#1}}%
                       \renewcommand{\stdlabel}{#1}%
                       #2%
                       \renewcommand{\Maximum}{}%
                      }%
                }%
               {\renewcommand{\@currentlabel}{\Maximum}%
                \label{listnr\arabic{listennr}}%
                \end{list}%
                }%
\makeatother
\newenvironment{StandardEinrueckung}[2]%
               {\begin{list}%
                      {#1}%
                      {\settowidth{\leftmargin}{M#1}%
                       \settowidth{\labelwidth}{#1}%
                       \settowidth{\labelsep}{M}%
                       #2%
                      }%
                \item}%
               {\end{list}}%
\newenvironment{Einrueckungpur}[1]%
               {\begin{StandardEinrueckung}{#1}{\ListNullAbstaende}}%
               {\end{StandardEinrueckung}}%
\newenvironment{Einrueckung}[1]%
               {\begin{StandardEinrueckung}{#1}{\setlength{\parsep}{0pt}}}%
               {\end{StandardEinrueckung}}%
\newcommand{\EineZeileGleichung}[2][0.0ex]
           {
            
            \vspace{#1} 
            \noindent
            \hspace*{\fill}
            $\displaystyle{#2}$
            \hspace*{\fill}

            \vspace{#1} 
            
           }

\makeatletter
\newcommand{\EineNumZeileGleichung}[2][0.5ex]
           {
            
            \vspace{#1} 
            \noindent
            \stepcounter{equation}
            \renewcommand{\@currentlabel}{\arabic{equation}}%
            \phantom{(\arabic{equation})}\hspace*{\fill}
            $\displaystyle{#2}$
            \hspace*{\fill}
            (\arabic{equation})

            \vspace{#1} 
            
           }
\makeatother
\makeatletter
\newcommand{\EineErwNumZeileGleichung}[2][0.5ex]
           {
            
            \vspace{#1} 
            \noindent
            \stepcounter{equation}
            \renewcommand{\@currentlabel}{\arabic{equation}}%
            \phantom{(\arabic{equation})}\hspace*{\fill}
            #2 %
            \hspace*{\fill}
            (\arabic{equation})

            \vspace{#1} 
            
           }
\makeatother
\newcommand{\breitrel}[1]{\hspace*{\tabcolsep} #1 \hspace*{\tabcolsep}}
\newlength{\abstaug}              %
\newenvironment{AllgUnnumGleichung}[2][1.0ex]
               {
  
                \setlength{\abstaug}{#1}
                \vspace{\abstaug}
                \hspace*{\fill}
                $\begin{array}[t]{#2}
                }%
               {\end{array}$
                \hspace*{\fill}
  
                \vspace{\abstaug}

                }%
\newenvironment{AllgNumGleichung}[2][0.0ex]
               {
  
                \setlength{\abstaug}{#1}
                \vspace{\abstaug}
                $\begin{tabular*}{\textwidth}[t]{#2}
                }%
               {\end{tabular*}$

                \vspace{\abstaug}

               }%
\newenvironment{StandardUnnumGleichungKlein}[1][0ex]
               {\renewcommand{\s}{\\[#1] }%
                \begin{AllgUnnumGleichung}{rcl}}%
               {\end{AllgUnnumGleichung}}%
\newenvironment{StandardUnnumGleichung}[1][0ex]%
               {\renewcommand{\s}{\\[#1] }%
                \begin{AllgUnnumGleichung}{>{\displaystyle}rc>{\displaystyle}l}}%
               {\end{AllgUnnumGleichung}}%
\newenvironment{XrelYZNumGleichung}[1][0ex]
               {\renewcommand{\s}{\\[#1] }%
                \begin{AllgNumGleichung}{rcll}}%
               {\end{AllgNumGleichung}}%
\newcommand{\erl}[1]{\hfill\mbox{\hspace*{1.5em}\small (#1)}}

\newcommand{\erllang}[2][0.5\textwidth]%
              {\hfill\hspace*{1.5em}%
               \begin{minipage}[t]{#1}{\small%
                          \begin{list}{(}{\ListNullAbstaende%
                                          \settowidth{\leftmargin}{(}%
                                          \settowidth{\labelwidth}{(}%
                                          \settowidth{\labelsep}{}%
                                         }%
                          \item#2)%
                          \end{list}}%
               \end{minipage}\\[-0.9ex]
              }%
\newcommand{\DefBemUmgeb}[1]%
           {\newenvironment{#1}[1][]%
                           {\begin{Einrueckung}{{\bf #1}}%
                            \ifx##1\empty\else{{\bf ##1}
                            
                                                        }\fi%
                            }%
                           {\end{Einrueckung}}}
\newcommand{\DefSBemUmgeb}[2]
           {\newenvironment{#1}[1][]%
                           {\begin{Einrueckung}{{\bf #2}}%
                            \ifx##1\empty\else{{\bf ##1}
                            
                                                        }\fi%
                            }%
                           {\end{Einrueckung}}}
\makeatletter
\newcommand{\DefBspUmgeb}[3]
           {\newcounter{#2}[#3]%
            \newenvironment{#1}[1][]%
                           {\stepcounter{#2}%
                            \renewcommand{\ZaehlerMarke}{\arabic{#2}}%
                            \renewcommand{\Einzugsname}{{\bf #1 \ZaehlerMarke}}%
                            \begin{Einrueckung}{\Einzugsname}
                            \ifx##1\empty\else{{\bf ##1}\\}\fi%
                            \renewcommand{\@currentlabel}{\ZaehlerMarke}%
                            }%
                           {\end{Einrueckung}}}
\makeatother
\newcommand{\ZaehlerbisEbene}{section}
\newcommand{\Ebenea}{section}
\newcommand{\Ebeneb}{subsection}

\newcommand{\Abschnittnummer}{%
            \ifx\ZaehlerbisEbene\Ebenea{\arabic{section}}%
             \else{%
              \ifx\ZaehlerbisEbene\Ebeneb{\arabic{section}.\arabic{subsection}}%
               \else{\arabic{section}.\arabic{subsection}.\arabic{subsubsection}}%
              \fi}%
            \fi}     
\newcommand{\Abschnittnummerpunkt}{\Abschnittnummer.}     
\newcommand{\Einzugsname}{}
\newcommand{\ZaehlerMarke}{}
\makeatletter
\newcommand{\DefThmUmgeb}[3]%
           {\newcounter{#1}[#3]%
            \newenvironment{#1}[1][]%
                           {\stepcounter{#2}%
                            \setcounter{#1}{\value{#2}}%
                            \renewcommand{\ZaehlerMarke}{\Abschnittnummerpunkt\arabic{#1}}%
                            \renewcommand{\Einzugsname}{{\bf #1 \ZaehlerMarke}}%
                            \begin{Einrueckung}{\Einzugsname}
                            \ifx##1\empty\else{{\bf ##1}
                            
                                                        }\fi%
                            \renewcommand{\@currentlabel}{\ZaehlerMarke}%
                            }%
                           {\end{Einrueckung}}}
\makeatother
\makeatletter
\newcommand{\DefSThmUmgeb}[4]%
           {\newcounter{#1}[#3]%
            \newenvironment{#1}[1][]%
                           {\stepcounter{#2}%
                            \setcounter{#1}{\value{#2}}%
                            \renewcommand{\ZaehlerMarke}{\Abschnittnummerpunkt\arabic{#1}}%
                            \renewcommand{\Einzugsname}{{\bf #4 \ZaehlerMarke}}
                            \begin{Einrueckung}{\Einzugsname}
                            \ifx##1\empty\else{{\bf ##1}

                                                        }\fi%
                            \renewcommand{\@currentlabel}{\ZaehlerMarke}%
                            }%
                           {\end{Einrueckung}}}
\makeatother
\makeatletter
\newcommand{\DefUnterNumThmUmgeb}[5]%
           {\newcounter{#1}[#3]%
            \newcounter{#4}%
            \newenvironment{#1}[1][]%
                           {\ifx##1\empty\else{\stepcounter{#2}\setcounter{#4}{0}}\fi%
                            \stepcounter{#4}%
                            \setcounter{#1}{\value{#2}}%
                            \renewcommand{\ZaehlerMarke}{\Abschnittnummerpunkt\arabic{#1}\alph{#4}}%
                            \renewcommand{\Einzugsname}{{\bf #5 \ZaehlerMarke}}
                            \begin{Einrueckung}{\Einzugsname}
                            \renewcommand{\@currentlabel}{\ZaehlerMarke}%
                            }%
                           {\end{Einrueckung}}}
\makeatother
\newenvironment{Beweis}[1][]%
               {\begin{Einrueckung}{{\bf Beweis}}%
                \ifx#1\empty\else{{\bf #1}

                                            }\fi%
                }%
               {\end{Einrueckung}%
                }%
\newenvironment{Proof}[1][]%
               {\begin{Einrueckung}{{\bf Proof}}%
                \ifx#1\empty\else{{\bf #1}

                                            }\fi%
                }%
               {\end{Einrueckung}%
                }%
               {\begin{Einrueckung}{{\bf \glqq Beweis\grqq}}%
                \ifx#1\empty\else{{\bf #1}
                
                                            }\fi%
                }%
               {\end{Einrueckung}%
                }%
               {\begin{Einrueckung}{{\bf Begr"undung}}%
                \ifx#1\empty\else{{\bf #1}
                
                                            }\fi%
                }%
               {\end{Einrueckung}%
                }%
\newenvironment{Hinrichtung}%
               {\begin{Einrueckungpur}{$\impliz$}}%
               {\end{Einrueckungpur}}%
\newenvironment{Rueckrichtung}%
               {\begin{Einrueckungpur}{$\invimpliz$}}%
               {\end{Einrueckungpur}}%
               {\begin{Einrueckungpur}{\glqq$\teilmenge$\grqq}}%
               {\end{Einrueckungpur}}%
               {\begin{Einrueckungpur}{\glqq$\obermenge$\grqq}}%
               {\end{Einrueckungpur}}%
               {\begin{Einrueckungpur}{"$\teilmenge$"}}%
               {\end{Einrueckungpur}}%
               {\begin{Einrueckungpur}{"$\obermenge$"}}%
               {\end{Einrueckungpur}}%
\newcommand{\qed}{\nopagebreak\hspace*{2em}\hspace*{\fill}{\bf qed}}
\newcommand{\ARabic}{\arabic}
\newcommand{\Nummerntypa}{\arabic}   
\newcommand{\Nummerntypb}{\alph}
\newcommand{\Nummerntypc}{\roman}
\newcommand{\Nummerntypd}{\Alph}

\newcommand{\Nra}{\Nummerntypa{Nummera}}            %
\newcommand{\Nrb}{\Nummerntypb{Nummerb}}            %
\newcommand{\Nrc}{\Nummerntypc{Nummerc}}                
\newcommand{\Nrd}{\Nummerntypd{Nummerd}}                
\newcommand{\ZeichenzuNrTyp}[1]%
           {\ifx#1\ARabic {.}\else{)}%
                  \fi}                              %
\newcommand{\NrZeicha}{\ZeichenzuNrTyp{\Nummerntypa}}
\newcommand{\NrZeichb}{\ZeichenzuNrTyp{\Nummerntypb}}
\newcommand{\NrZeichc}{\ZeichenzuNrTyp{\Nummerntypc}}
\newcommand{\NrZeichd}{\ZeichenzuNrTyp{\Nummerntypd}}
\newcommand{\ListMarkea}%
           {\Nra\NrZeicha}
\newcommand{\ListMarkeb}%
           {\Nra\NrZeicha\Nrb\NrZeichb}
\newcommand{\ListMarkec}%
           {\Nra\NrZeicha\Nrb\NrZeichb\Nrc\NrZeichc}
\newcommand{\ListMarked}%
           {\Nra\NrZeicha\Nrb\NrZeichb\Nrc\NrZeichc\Nrd\NrZeichd}
\newcommand{\Anfangszeichen}{}
\newcommand{\Anfangspunkt}{}
\newcounter{Schachtelebene}
\newcounter{Hilfszaehler}
\newcommand{\Hilfsbefehl}{}
\newcommand{\Schachtelebene}{\alph{Schachtelebene}}
\makeatletter
\newenvironment{AllgNumerierteListe}[2][]
               {\addtocounter{Schachtelebene}{1}%
		\setcounter{Hilfszaehler}{#2}%
                \renewcommand{\Anfangszeichen}%
                             {\renewcommand{\Hilfsbefehl}{\csname Nummerntyp\Schachtelebene \endcsname}%
                              \Hilfsbefehl{Hilfszaehler}}%
                \renewcommand{\Anfangspunkt}%
                             {\csname NrZeich\Schachtelebene \endcsname}%
                \begin{list}%
                      {\stepcounter{Nummer\Schachtelebene}%
                       \csname Nr\Schachtelebene \endcsname
                       \csname NrZeich\Schachtelebene \endcsname
                       }%
                      {\settowidth{\leftmargin}{M\Anfangszeichen\Anfangspunkt}%
                       \settowidth{\labelwidth}{\Anfangszeichen\Anfangspunkt}%
                       \settowidth{\labelsep}{M}%
                       \setlength{\topsep}{0pt}%
                       \setlength{\parskip}{0pt}%
                       \setlength{\partopsep}{0pt}%
                       \setlength{\itemsep}{0pt}%
                       \setlength{\parsep}{0pt}%
                      }%
                \renewcommand{\@currentlabel}{\csname ListMarke\Schachtelebene \endcsname}%
                }%
               {\ifthenelse{\equal{}{}}{\setcounter{Nummer\Schachtelebene}{0}}{}
                \addtocounter{Schachtelebene}{-1}%
                \end{list}}
\makeatother
\newenvironment{NumerierteListe}[1]
               {\begin{AllgNumerierteListe}{#1}}
               {\end{AllgNumerierteListe}}
\newenvironment{WeiterNumerierteListe}[1]
               {\begin{AllgNumerierteListe}[Weiter]{#1}}
               {\end{AllgNumerierteListe}}

\newcommand{\UnnumAnfangszeichen}{}
\newcounter{UnnumSchachtelebene}
\newcommand{\UnnumSchachtelebene}{\alph{UnnumSchachtelebene}}
\makeatletter
\newenvironment{UnnumerierteListe}%
               {\addtocounter{UnnumSchachtelebene}{1}%
                \renewcommand{\UnnumAnfangszeichen}%
                             {\csname UnnumZeich\UnnumSchachtelebene \endcsname}%
                \begin{list}%
                      {\UnnumAnfangszeichen}%
                      {\settowidth{\leftmargin}{M\UnnumAnfangszeichen}%
                       \settowidth{\labelwidth}{\UnnumAnfangszeichen}%
                       \settowidth{\labelsep}{M}%
                       \setlength{\topsep}{0pt}%
                       \setlength{\parskip}{0pt}%
                       \setlength{\partopsep}{0pt}%
                       \setlength{\itemsep}{0pt}%
                       \setlength{\parsep}{0pt}%
                      }%
                }%
               {\addtocounter{UnnumSchachtelebene}{-1}%
                \end{list}}
\makeatother
\newlength{\fktdefhilfslaenge}
\newcommand{\ohnefktdef}[4]
           {\hspace*{\fill}
            $\begin{array}[t]{ccc}%
            #1 & \nach & #2 \\
            #3 & \auf  & #4
            \end{array}$
            \hspace*{\fill}}
\newcommand{\fktdef}[5]
           {\hspace*{\fill}
            $\begin{array}[t]{cccc}%
            #1: & #2 & \nach & #3 \\    
                & #4 & \auf  & #5
            \end{array}$
            \settowidth{\fktdefhilfslaenge}{$#1$:}
            \hspace*{0.6 \fktdefhilfslaenge}  
            \hspace*{\fill}}
\newcommand{\fktdefpur}[5]
           {$\begin{array}[t]{cccc}%
            #1: & #2 & \nach & #3 \\    
                & #4 & \auf  & #5
            \end{array}$}
\newcommand{\fktdefabgesetztpur}[5]
           {
            
            $\begin{array}[t]{cccc}%
            #1: & #2 & \nach & #3 \\    
                & #4 & \auf  & #5
            \end{array}$
            \settowidth{\fktdefhilfslaenge}{$#1$:}
            \hspace*{0.6 \fktdefhilfslaenge}
            
           }
\newcommand{\fktdefabgesetzt}[5]
           {
           
            \hspace*{\fill}
            $\begin{array}[t]{cccc}%
            #1: & #2 & \nach & #3 \\    
                & #4 & \auf  & #5
            \end{array}$
            \settowidth{\fktdefhilfslaenge}{$#1$:}
            \hspace*{0.6 \fktdefhilfslaenge}  
            \hspace*{\fill}
            
            }
\newcommand{\ohnefktdefabgesetzt}[4]
           {      

            \hspace*{\fill}
            $\begin{array}[t]{ccc}%
            #1 & \nach & #2 \\
            #3 & \auf  & #4
            \end{array}$
            \hspace*{\fill}

            }
\newcommand{\doppelohnefktdefabgesetzt}[6]
           {

            \hspace*{\fill}
            $\begin{array}[t]{ccccc}%
            #1 & \nach & #2 & \nach & #3\\
            #4 & \auf  & #5 & \auf  & #6
            \end{array}$
            \hspace*{\fill}

            }
\newcommand{\anhang}%
           {\appendix
            \sectioninh{Anhang}
            \renewcommand{\Abschnittnummer}{%
                  \ifx\ZaehlerbisEbene\Ebenea{\Alph{section}}%
                  \else{%
                        \ifx\ZaehlerbisEbene\Ebeneb{\Alph{section}.\arabic{subsection}}%
                        \else{\Alph{section}.\arabic{subsection}.\arabic{subsubsection}}%
                        \fi}%
                  \fi}%
            \renewcommand{\Abschnittnummerpunkt}{\Abschnittnummer.}     
            }            
\newcommand{\anhangengl}%
           {\appendix
            \sectioninh{Appendix}
            \renewcommand{\Abschnittnummer}{%
                  \ifx\ZaehlerbisEbene\Ebenea{\Alph{section}}%
                  \else{%
                        \ifx\ZaehlerbisEbene\Ebeneb{\Alph{section}.\arabic{subsection}}%
                        \else{\Alph{section}.\arabic{subsection}.\arabic{subsubsection}}%
                        \fi}%
                  \fi}%
            \renewcommand{\Abschnittnummerpunkt}{\Abschnittnummer.}     
            }

\newcounter{wdhlstufe}
\newcommand{\sectioninh}[1]%
           {\section*{#1}%
            \addcontentsline{toc}{section}{#1}}
\newcommand{\bezeichnung}[3]%
           {\begin{Einrueckungpur}{\hbox to 6em{#1}\hbox to 2.4em{\hfill#2}}
            #3
            \end{Einrueckungpur}}

\newcommand{\doppelteinfach}{e}

\newcommand{\ifdoppelt}[1]{\ifthenelse{\equal{\doppelteinfach}{d}}{#1}{}}
\newcommand{\ifeinfach}[1]{\ifthenelse{\equal{\doppelteinfach}{e}}{#1}{}}

\newlength{\querfhilfsl}              %

\newlength{\hll}

\DefThmUmgeb{Theorem}{Theorem}{\ZaehlerbisEbene}
\DefThmUmgeb{Definition}{Definition}{\ZaehlerbisEbene}
\DefThmUmgeb{Satz}{Theorem}{\ZaehlerbisEbene}
\DefThmUmgeb{Proposition}{Theorem}{\ZaehlerbisEbene}
\DefThmUmgeb{Lemma}{Theorem}{\ZaehlerbisEbene}
\DefThmUmgeb{Folgerung}{Theorem}{\ZaehlerbisEbene}
\DefThmUmgeb{Corollary}{Theorem}{\ZaehlerbisEbene}
\DefThmUmgeb{Vorschrift}{Definition}{\ZaehlerbisEbene}
\DefThmUmgeb{Construction}{Definition}{\ZaehlerbisEbene}
\DefSThmUmgeb{FormSatz}{Theorem}{\ZaehlerbisEbene}{\glqq Satz\grqq} 
\DefThmUmgeb{Vermutung}{Theorem}{\ZaehlerbisEbene}
\DefThmUmgeb{Conjecture}{Theorem}{\ZaehlerbisEbene}
\DefThmUmgeb{Konvention}{Definition}{\ZaehlerbisEbene}
\DefThmUmgeb{Feststellung}{Theorem}{\ZaehlerbisEbene}
\DefUnterNumThmUmgeb{DefinitionZusatzNum}{Definition}{\ZaehlerbisEbene}{DefZN}{Definition}
\DefBspUmgeb{Beispiel}{Beispiel}{subsubsection}
\DefBspUmgeb{Example}{Example}{subsubsection}
\DefBspUmgeb{Frage}{Frage}{section}
\DefBspUmgeb{Question}{Question}{section}
\DefBspUmgeb{Aufgabe}{Aufgabe}{section}
\DefBspUmgeb{Ziel}{Ziel}{section}
\DefBemUmgeb{Bemerkung}
\DefBemUmgeb{Remark}
\DefSBemUmgeb{OffeneFrage}{Offene Frage}
\newcommand{\bdf}{\begin{Definition}}
\newcommand{\edf}{\end{Definition}}
\newcommand{\bvorsch}{\begin{Vorschrift}}
\newcommand{\evorsch}{\end{Vorschrift}}
\newcommand{\bconst}{\begin{Construction}}
\newcommand{\econst}{\end{Construction}}
\newcommand{\bthm}{\begin{Theorem}}
\newcommand{\ethm}{\end{Theorem}}
\newcommand{\bsatz}{\begin{Satz}}
\newcommand{\esatz}{\end{Satz}}
\newcommand{\bprop}{\begin{Proposition}}
\newcommand{\eprop}{\end{Proposition}}
\newcommand{\blem}{\begin{Lemma}}
\newcommand{\elem}{\end{Lemma}}
\newcommand{\bfolg}{\begin{Folgerung}}
\newcommand{\efolg}{\end{Folgerung}}
\newcommand{\bcorr}{\begin{Corollary}}
\newcommand{\ecorr}{\end{Corollary}}
\newcommand{\bfest}{\begin{Feststellung}}
\newcommand{\efest}{\end{Feststellung}}
\newcommand{\bbew}{\begin{Beweis}}
\newcommand{\ebew}{\end{Beweis}}
\newcommand{\bpf}{\begin{Proof}}
\newcommand{\epf}{\end{Proof}}
\newcommand{\bwnum}{\begin{WeiterNumerierteListe}}
\newcommand{\ewnum}{\end{WeiterNumerierteListe}}
\newcommand{\bdfzn}{\begin{DefinitionZusatzNum}}
\newcommand{\edfzn}{\end{DefinitionZusatzNum}}
\newcommand{\bbem}{\begin{Bemerkung}}
\newcommand{\ebem}{\end{Bemerkung}}
\newcommand{\brem}{\begin{Remark}}
\newcommand{\erem}{\end{Remark}}
\newcommand{\bnum}{\begin{NumerierteListe}}
\newcommand{\enum}{\end{NumerierteListe}}
\newcommand{\bunum}{\begin{UnnumerierteListe}}
\newcommand{\eunum}{\end{UnnumerierteListe}}
\newcommand{\bbsp}{\begin{Beispiel}}
\newcommand{\ebsp}{\end{Beispiel}}
\newcommand{\bex}{\begin{Example}}
\newcommand{\eex}{\end{Example}}
\newcommand{\bfrag}{\begin{Frage}}
\newcommand{\efrag}{\end{Frage}}
\newcommand{\bquest}{\begin{Question}}
\newcommand{\equest}{\end{Question}}
\newcommand{\baufg}{\begin{Aufgabe}}
\newcommand{\eaufg}{\end{Aufgabe}}
\newcommand{\bof}{\begin{OffeneFrage}}
\newcommand{\eof}{\end{OffeneFrage}}
\newcommand{\bverm}{\begin{Vermutung}}
\newcommand{\everm}{\end{Vermutung}}
\newcommand{\bconj}{\begin{Conjecture}}
\newcommand{\econj}{\end{Conjecture}}
\newcommand{\bkonv}{\begin{Konvention}}
\newcommand{\ekonv}{\end{Konvention}}
\newcommand{\bglklein}{\begin{StandardUnnumGleichungKlein}}
\newcommand{\eglklein}{\end{StandardUnnumGleichungKlein}}
\newcommand{\bgl}{\begin{StandardUnnumGleichung}}
\newcommand{\egl}{\end{StandardUnnumGleichung}}
\newcommand{\bglrtext}{\begin{XrelYZNumGleichung}}
\newcommand{\eglrtext}{\end{XrelYZNumGleichung}}
\newcommand{\zgl}{\EineZeileGleichung}

\newcommand{\zglklein}[1]{\zgl{\textstyle#1}}

\newcommand{\berlgl}{\begin{StandardUnnumGleichung}}
\newcommand{\eerlgl}{\end{StandardUnnumGleichung}}
\newcommand{\beinrueck}{\begin{Einrueckungpur}} 
\newcommand{\eeinrueck}{\end{Einrueckungpur}}
\newcommand{\beinflist}{\begin{EinfachListe}} 
\newcommand{\eeinflist}{\end{EinfachListe}}
\newcommand{\beq}{\begin{equation}}
\newcommand{\eeq}{\end{equation}}
\newcommand{\bhin}{\begin{Hinrichtung}}
\newcommand{\ehin}{\end{Hinrichtung}}
\newcommand{\brueck}{\begin{Rueckrichtung}}
\newcommand{\erueck}{\end{Rueckrichtung}}
\newcommand{\bvl}{\begin{AutoLabelLaengenListe}{\ListNullAbstaende}}
\newcommand{\evl}{\end{AutoLabelLaengenListe}}
\newcommand{\df}[1]{{\bf #1}}

\renewcommand{\he}{{\text{ss}}}                      
\newcommand{\defi}[1]{\delta_{#1}}
\newcommand{\gf}{\gross f}

\newcommand{\cxy}[1]{{\cal V}_{#1}}
\newcommand{\was}{{\cal V}}            

\newcommand{\reg}{{\mathrm{reg}}}

\newcommand{\triv}{\iota}
\newcommand{\trv}[2][]{\ifthenelse{\equal{#1}{}}{{#2}^\triv}{#2^{#1}}}
 
\newcommand{\tass}{T}
\newcommand{\LF}{{\cal L}}              
\newcommand{\awpt}{H}

\newcommand{\pot}[3]{[#1_{#2}]^{\bullet #3}}

\newcommand{\web}{w}

\renewcommand{\he}{{\text{ss}}}

\newcommand{\cl}{\mathrm{cl}}

\newlength{\adressabstand}
\begin{document}
\title{Parallel Transports in Webs}
\author{Christian Fleischhack\thanks{e-mail: 
            {\tt chfl@mis.mpg.de}} \\   
        \\
        {\normalsize\em Max-Planck-Institut f\"ur Mathematik in den
                        Naturwissenschaften}\\[\adressabstand]
        {\normalsize\em Inselstra\ss e 22--26}\\[\adressabstand]
        {\normalsize\em 04103 Leipzig, Germany}
        \\[-25\adressabstand]      
        {\normalsize\em Center for Gravitational Physics and Geometry}\\[\adressabstand]
        {\normalsize\em 320 Osmond Lab}\\[\adressabstand]
        {\normalsize\em Penn State University}\\[\adressabstand]
        {\normalsize\em University Park, PA 16802}
        \\[-25\adressabstand]}      
\date{July 17, 2003}
\maketitle
\begin{abstract}
For connected reductive linear algebraic structure groups it is proven that 
every web is holonomically isolated. The possible tuples of parallel transports 
in a web form a Lie subgroup of the corresponding power
of the structure group. This Lie subgroup is explicitly calculated 
and turns out to be independent of the chosen local trivializations. 
Moreover, explicit necessary and sufficient 
criteria for the holonomical independence of webs are derived.
The results above can even be sharpened: 
Given an arbitrary neighbourhood of the base points of a web,
then this neighbourhood contains some segments of the web 
whose parameter intervals coincide, but do not include 0 
(that corresponds to the base points of the web), and whose parallel transports
already form the same Lie subgroup as those of the full web do.
\end{abstract}

\section{Introduction}
In order to incorporate the full diffeomorphism invariance of general
relativity into loop gravity, it is not sufficient to consider only
piecewise analytic paths. Instead, at least, piecewise smooth and immersive
paths have to be included. This, however, causes a bunch of technical 
difficulties that are usually related to the fact that two finite graphs need
not be contained in a common larger finite graph. This desirable, but not given
directedness is necessary to, in particular, make the measure theory 
well defined. Several attempts have been made to circumvent this problem.
First, Baez and Sawin \cite{d3} introduced so-called webs. These
are certain sets of piecewise smooth immersive paths that are 
``sufficiently'' independent to allow for the definition of the 
Ashtekar-Lewandowski measure. Later, arbitrarily smooth paths have been
shown tractable using hyphs \cite{paper3,diss}. Nevertheless, there are
still several difficulties even in the more restrictive case of smooth webs.
One of them is related to the Lewandowski-Thiemann conjecture \cite{e46},
which is important for the definition of diffeomorphism-invariant 
operators in loop quantum gravity. More precisely, since the averaging
over the diffeomorphism group is well defined only on a certain
subspace of cylindrical functions, it is important that diffeomorphism-invariant
operators do indeed preserve that subspace. Lewandowski and Thiemann
took the view that this is probably true and argued that to answer
that question one should study how parallel transports in portions of webs
behave. In the present article we are now going to state several 
results on this subject. In particular, we will be able to prove
that in the case of semisimple structure groups 
(like $SU(2)$ or $Sl(2,\C)$ needed for gravity) 
there are always subpaths of a web that do not run through the base points
of the web, but are already holonomically independent,
i.e.\ parallel transports can be assigned to them completely independent
of each other. However, one can see \cite{paper15} that these results are
--~in contrast to the anticipation expressed in \cite{e46}~-- 
still not completely sufficient to prove the Lewandowski-Thiemann conjecture.
Nevertheless, the methods developed here will be used to prove it
in a subsequent article \cite{paper15}.

The present paper goes as follows:
After some preliminaries
we will introduce the terms ``richness'' and ``splitting''. 
They will be used
to encode the relative position of (parts of) webs -- do they coincide,
are they in a certain sense independent? Next, we will partially 
exploit an idea, already used in \cite{e46}, to study under which circumstances
groups can be generated by finite products of elements of certain subgroups.
Together with some criteria for the holonomical independence
of sets of paths, we will finally determine for every web explicitly 
what parallel transports regular connections may
have. In particular, given a web of $n$ paths, one sees that the set of possible
parallel transports forms a Lie subgroup of $\LG^n$ for structure groups
$\LG$ that are compact Lie. This can be regarded as a proof that is 
independent from the argumentation in \cite{d3}. But, our result
is true even if $\LG$ is an arbitrary product of semisimple and abelian,
possibly noncompact Lie groups. Additionally, we will show that 
all parallel transports, occurring in a full web, can already 
be adopted along certain subpaths in a web that are nontrivial in the sense  
that they do not contain the base points of the web.

\section{Preliminaries}
Let us briefly fix the notations. Throughout the whole paper, 
$\LG$ is some arbitrary group. Starting with Section \ref{sect:holo_indep}
we assume additionally that $\LG$ is a connected Lie group.
Fix some arbitrary manifold $M$.
Let $\Pf$ denote the set of all (finite) paths in $M$, i.e.\ the set of all
piecewise smooth and immersive mappings from $[0,1]$ to $M$. 
$\Pf$ is a groupoid 
(after imposing the standard equivalence relation, i.e., saying that
reparametrizations and insertions/deletions of retracings are irrelevant) 
\cite{paper2+4,d3}.
Sometimes we will speak about paths restricted to certain
subintervals $I$ of $[0,1]$. By means of some affine map from $I$ to $[0,1]$
we may regard these restrictions naturally as paths again.
The set of all smooth connections in some fixed principal fibre bundle 
$\pi:P\nach M$ with structure group $\LG$ is denoted by $\A$.
Given some ``ultralocal'' trivialization $\triv$ of $P$, 
we can identify for every path $\gamma \in \Pf$ and every connection $A \in \A$
the parallel transport w.r.t.\ to $A$ along $\gamma$
with an element $\trv h_A(\gamma) \ident \trv h_\gamma(A)$ of $\LG$.
Moreover, we define for every finite tuple 
$\gc = (\gamma_1,\ldots,\gamma_n)$ of paths the set 
\zgl{
\trv\A_\gc 
 :=  \{\trv h_A(\gc)\mid A\in\A \} 
 \ident \bigl\{\bigl(\trv h_A(\gamma_1),\ldots,\trv h_A(\gamma_n)\bigr) 
               \mid A\in\A \bigr\} 
 \teilmenge \LG^n
}\noindent
of all possible (tuples of) parallel transports along these paths.
Recall that an ultralocal trivialization \cite{paper10} is simply 
some collection of trivializations
for each single fibre $P_m$ in $P$. Note that the assignment of $m\in M$ to
that trivialization over $m$ -- even locally -- need not be smooth.
Sometimes, however,
we will drop the superscript $\triv$ to simplify notation.
Then we assume we are given some arbitrary, but fixed trivialization.
It is obvious that $\trv\A_\gc$ is independent of the chosen
trivialization, in particular, if it equals the full $\LG^{\elanz\gc}$.

\section{Richness and Splittings}
\label{sect:richsplit}
Let $n\in\N_+$ be some positive integer and let $\LG$ be some group.

\bdf
\label{def:V_n}
We define
\bunum
\item
$\cxy n$ to be the set of all 
$n$-tuples with entries equal to $0$ or $1$ only;
\item
$\LG_v := \{(g^{v_1},\ldots,g^{v_n}) \mid g\in\LG\} \teilmenge \LG^n$
for every $v \in \cxy n$;      
and
\item
$\LG_V := \LG_{v^1} \:\cdots\: \LG_{v^k}$
for every ordered\footnote{By an ordered subset of $X$ we mean an arbitrary
tuple of elements in $X$ where every element in $X$ occurs at most once as
a component of that tuple. However, we will use the standard terminology 
of sets if misunderstandings seem to be impossible.}
subset $V = \{v^1,\ldots,v^k\} \teilmenge \cxy n$.
\eunum
\edf
We have, e.g., $\LG_{(1,0,1,0)} = \{(g,1,g,1) \mid g\in\LG\}$.

\blem
\label{lem:type=LieUGru}
For every $n\in\N_+$, every group (Lie group, algebraic group) $\LG$ 
and every $v\in \cxy n$,
the set $\LG_v$ is a subgroup (Lie subgroup, algebraic subgroup) 
of $\LG^n$. 
\elem
\bpf
Obviously, $\LG_v$ is a subgroup of $\LG$. If $\LG$ is Lie, then it
is additionally a submanifold, hence a Lie subgroup of $\LG^n$.
If $\LG$ is algebraic, then $\LG_v$ is Zariski-closed in $\LG$,
hence an algebraic subgroup.
\qed
\epf

\subsection{Richness}
\bdf
\label{def:rich}
An ordered subset $V\teilmenge \cxy n$ is called \df{rich} iff
\bnum{2}
\item
for all $1 \leq i,j \leq n$ with $i\neq j$ there is an element $v\in V$ with $v_i \neq v_j$ and
\item
for all $1 \leq i \leq n$ there is an element $w\in V$ with $w_i \neq 0$.
\enum
\edf
For instance, let $n=4$. 
Then $V := \{(1,1,0,0),\:(1,0,1,0),\:(0,1,0,1),\:(0,0,1,1)\}$ 
is rich, but $\{(1,1,0,1),\:(1,0,1,1),\:(0,1,1,0)\}$ is not, because it
fails to fulfill the first richness condition for $i=1$ and $j=4$. 
\bdf
Let $n \in \N_+$ and $K \echteteilmenge \{1,\ldots,n\}$. 
\bunum
\item
For every $v\in\cxy n$ the \df{$K$-restriction} $R_K(v) \in \cxy{n-\betrag K}$ 
of $v$ is defined to be the $(n - \betrag K)$-tuple, that is generated
from $v$ by canceling all components of $v$ at the positions listed in $K$.
\item
For every $V\teilmenge\cxy n$ the $K$-restriction of $V$ is given by 
\zglklein{R_K(V) := \bigcup_{v\in V} \{R_K(v)\}.}
\item
For every nonempty $V\teilmenge\cxy n$ the
\df{richness deficit} $\defi V$ of $V$ is given 
for $V \neq \{(0,\ldots,0)\}$ by
\zgl{\defi V := \min_{K' \echteteilmenge \{1,\ldots,n\}} 
              \{\betrag{K'}\in\N \mid \text{$R_{K'}(V)$ is rich}\}}
and by $\defi V := n$ otherwise.
\eunum
\edf
For example, we have $R_{\{2,4\}}\bigl((1,0,1,0)\bigr) = (1,1)$ and 
$R_{1}\bigl((1,0,1,0)\bigr) = (0,1,0)$.
For convenience, we write $R_k$ instead of $R_{\{k\}}$ for $1\leq k \leq n$.
Using the example above of a rich $V$, we get
$R_4(V) = \{(1,1,0),\:(1,0,1),\:(0,1,0),\:(0,0,1)\}$.
Note, finally, that $\defi V$ is well defined. 
In fact, if $V$ is neither empty nor the zero tuple, it contains at least 
one nonzero tuple, say $v$ with $v_i = 1$. There
we have $R_{\{1,\ldots,n\} \setminus \{i\}}(V) = \{(1)\}$ which is rich.
\newpage
\blem
\label{lem:restrict+rich}
Let $V \teilmenge \cxy n$ be nonempty.
\bnum{3}
\item
If $V$ is rich, 
then $R_K(V)$ is rich for all $K \echteteilmenge \{1,\ldots,n\}$.
\item
If $(g_1,\ldots,g_{k-1},g_{k+1},\ldots,g_{n}) \in \pot\LG {R_k(V)} q$ for some
$q\in\N$, 
then there is some 
$g\in\LG$ with $(g_1,\ldots,g_{k-1},g,g_{k+1},\ldots,g_n) \in \pot\LG V q$.
\item
$V$ is rich iff $\defi V$ is zero.
\enum
\elem
\bpf
Clear.
\qed
\epf
We remark that $\pot\LG V q$ denotes the $q$-fold 
multiplication $\LG_V\cdots\LG_V$ of $\LG_V$. In contrast, we
use $\LG^n$ as usual for the $n$-fold direct product 
$\LG \kreuz \cdots \kreuz \LG$ of $\LG$.

\subsection{Splittings}

\bdf
\label{def:splitting}
\bunum
\item
A subset $V \teilmenge \cxy n$ is called 
\df{$n$-splitting} 
iff
\bnum{2}
\item
$\sum_{v\in V} v = (1,\ldots,1)$ and
\item
$(0,\ldots,0) \nichtin V$.
\enum
\item
Let $V$ and $V'$ be $n$-splittings. $V'$ is called \df{refinement}
of $V$ (shortly: $V' \geq V$) iff every $v \in V$ can be written
as a sum of elements in $V'$.
\eunum
\edf

\blem
\label{lem:split-gru-eig}
Let $V$ be some $n$-splitting. Then we have:
\bunum
\item
$\LG_V = \prod_{v\in V} \LG_v$ independently of the ordering in $V$;
\item
$\LG_V$ is a subgroup of $\LG^n$, hence $\pot\LG V 2 = \LG_V$;
\item
$\LG_{V'} \obermenge \LG_{V}$ for all $n$-splittings $V' \geq V$.
\eunum
\elem
\bpf
It is easy to see that $\LG_{v'}$ and $\LG_{v''}$ commute for 
all $v',v''\in \cxy n$ with $v' v'' = (0,\ldots,0)$ where the 
multiplication is pointwise. Moreover, we then have 
$\LG_{v'+v''} \teilmenge \LG_{v'}\LG_{v''}$.
Since, as follows directly from the definition, $v' v'' = (0,\ldots,0)$
for all different elements $v'$ and $v''$ in an $n$-splitting,
$\prod_{v\in V} \LG_v$ does not depend on the ordering. Now,
$\LG_V = \prod_{v\in V} \LG_v$ by definition. 
From $(\LG_v)^{-1} = \LG_v = \LG_v \LG_v$
for all $v\in\cxy n$, we get the group property of 
$\LG_V$ and hence $\pot\LG V 2 = \LG_V$.
Additionally, if $V' \geq V$, then every $v\in V$ is a sum of certain
elements in $V'$. Consequently, each $\LG_v$ is contained in $\LG_{V'}$,
whence $\LG_V \teilmenge \pot\LG{V'}{\elanz V} = \LG_{V'}$.
\qed
\epf

\bdf
\label{def:splitV(s)}
Let $n\in\N_+$ be some positive integer, $S$ be some set and
$\vec s$ be some $n$-tuple of elements of $S$.
Then the \df{splitting $V(\vec s)$ for $\vec s$} is given by
\zgl{V(\vec s) := 
     \{v\in \cxy n \mid 
      \text{$v_i = 1 = v_j$ $\aequ$ $s_i = s_j$}\}
      \setminus\{(0,\ldots,0)\}.}
\edf
For example, the splitting for $\vec s = (s_1,s_2,s_3,s_2)$
is $V(\vec s) = \{(1,0,0,0),(0,1,0,1),(0,0,1,0)\}$.

\blem
For every $n$, $S$ and $\vec s$ as given in Definition \ref{def:splitV(s)},
$V(\vec s)$ is a $n$-splitting.
\elem
\bpf
Let $1\leq i\leq n$ and $v,v'\in V(\vec s)$ with $v_i = v'_i = 1$. Then
$v_j = 1$ iff $s_i = s_j$. 
However, the same is true for $v'_j$. Hence, $v = v'$.
Since there is at least one $v\in \cxy n$ with $v_i = 1$, we get the assertion.
\qed
\epf

\bprop
\label{prop:uhalbstet(split)}
Let $\gf = (f_1,\ldots,f_n)$ be some tuple of continuous
functions $f_i : X \nach Y$, where $X$ and $Y$ are topological spaces and
$Y$ is assumed Hausdorff. 

Then for every $x_0\in X$ there
is some neighbourhood $U \teilmenge X$ of $x_0$,
such that $V(\gf(x_0)) \leq V(\gf(x))$ for all $x \in U$.
\eprop
\bpf
\bunum
\item
By the continuity and the finiteness of $\gf$, there is for every fixed
$x_0\in X$ some open $U \teilmenge X$ containing $x_0$, such that
$f_i(x_0) \neq f_j(x_0)$ implies $f_i(x) \neq f_j(x)$
for all $x\in U$. Note that $Y$ is Hausdorff.
\item
Set for every two $n$-splittings $V,V'$
\zgl{\kappa_{V'}(v) := \{v'\in V' \mid \esex k: v_k = 1 = v'_k\}.}
Then we have $v \leq \sum_{v' \in \kappa_{V'}(v)} v'$ 
(with $\leq$ defined by $\leq$ on all components). 
In fact, if $v_l = 0$, the assertion
$v_l \leq \sum_{v' \in \kappa_{V'}(v)} v'_l$
is trivial. On the other hand, for $v_l = 1$ 
there is some $\dach v'\in V'$ with 
$\dach v'_l = 1$, i.e.\ $\dach v' \in \kappa_{V'}(v)$, hence
$v_l = 1 \leq \sum_{v' \in \kappa_{V'}(v)} v'_l$.
\item
Observe now, that
$\kappa_{V(\gf(x))}(v) \cap \kappa_{V(\gf(x))}(\dach v) \neq \leeremenge$
with $x\in U$ and $v,\dach v\in V(\gf(x_0))$ implies $v = \dach v$. In fact,
it implies the existence of some $v'$ in that intersection
and some $k,l$ with $v'_k = 1 = v_k$ and $v'_l = 1 = \dach v_l$.
Consequently, $v'_k = 1 = v'_l$, hence $f_k(x) = f_l(x)$,
thus $f_k(x_0) = f_l(x_0)$ by $x \in U$. Therefore, $v_k = 1 = v_l$
and $\dach v_k = 1 = \dach v_l$ by $v, \dach v \in V(\gf(x_0))$.
This implies $v = \dach v$.
\item
Altogether we get for every $x\in U$
\bglklein
(1,\ldots,1) 
 &  =   & \sum_{v\in V(\gf(x_0))} v  
               \erl{Splitting property} \\
 & \leq & \sum_{v\in V(\gf(x_0))} \sum_{v'\in\kappa_{V(\gf(x))}(v)} v' 
               \erl{$v \leq \sum_{v' \in \kappa_{V(\gf(x))}(v)} v'$} \\
 & \leq & \sum_{v'\in V(\gf(x))} v' 
               \erl{Disjointness of the sets $\kappa_{V(\gf(x))}(v)$}\\
 &  =   & (1,\ldots,1) 
               \erl{Splitting property} 
\eglklein
and, therefore, $v = \sum_{v' \in \kappa_{V(\gf(x))}(v)} v'$ 
with $\kappa_{V(\gf(x))}(v) \teilmenge V(\gf(x))$ by definition
for every $v \in V(\gf(x_0))$. 
\qed
\eunum
\epf

\section{Generation of Subgroups}
Even if some smooth paths are independent, i.e.\ they are webs or hyphs,
two or more of them may share full segments.
Consequently, the parallel transports along these segments are identical.
Using the terminology of the previous section, we see that 
these parallel transports have now values in some subset $\LG_V$ of $\LG^n$,
where $V$ encodes which segments coincide. In order to prepare the study of
the behaviour of the parallel transports along the full paths,
we will now present some results on products of sets of type $\LG_V$.
\subsection{Semisimple Groups}
Before stating the first theorem of this paper, let us recall 
the definition of the commutator length \cite{m2}. 
\bdf
Let $\LG$ be a group that coincides with its commutator subgroup.
\bunum
\item
The \df{commutator length} $\cl_\LG(g)$ 
of an element $g \in \LG$ is defined to be 
the minimal number of commutators in $\LG$ whose product equals $g$.
\item
The \df{commutator length $\cl(\LG)$} of $\LG$ is defined to be 
$\sup_{g\in\LG} \cl_\LG(g)$.
\eunum
\edf
We remark that $\cl(\LG) = 1$ for all connected semisimple Lie groups
that are complex or compact \cite{m2}. This means, every element in $\LG$
can then be written as a commutator of two elements in $\LG$.
\bthm
\label{thm:rich->full}
Let $\LG$ be a group that equals its commutator subgroup
and let $n$ be some positive integer. Then we have
$\LG^n = \bigcup_{q\in\N} \: \pot\LG V q$ 
for any rich ordered subset $V$ of $\cxy n$.

If, moreover, $\LG$ has finite commutator length, then 
there is a positive integer $q(n)$ 
such that $\pot\LG V{q(n)} = \LG^n$ 
for any rich ordered subset $V$ of $\cxy n$.
\ethm
Note that $q(n)$ does not depend on the ordering or the number
of elements in $V$. 

The following proof will exploit an idea presented in \cite{e46}
for Lie algebras, but now we transfer it to the level of abstract groups.
\bpf
Let us first prove 
$\LG^n = \bigcup_{q\in\N} \: \pot\LG V q$.
\bunum
\item
The case $n=1$ is trivial.
\item
Let us consider the case $n=2$. By the first richness condition,
$V$ has to contain
at least one of the elements $(0,1)$ and $(1,0)$. W.l.o.g.\ we 
have $(0,1) \in V$. Then by the second richness condition, $(1,0)$ 
or $(1,1)$ is in $V$. However, we see immediately
that both for $i$ equal $0$ and $1$ we have 
$\LG^2 = \LG_{(0,1)} \LG_{(1,i)} \teilmenge \LG_V \teilmenge \LG^2$
if we assume that $(0,1)$ occurs in $V$ before $(1,i)$ does.
Completely analogously we have $\LG^2 = \LG_{(1,i)} \LG_{(0,1)} = \LG_V$ 
in the opposite case.
Hence, the assertion follows from
$\LG^2 = \LG_V \teilmenge \bigcup_{q\in\N} \: \pot\LG V q \teilmenge \LG^2$.
\item
We proceed by induction. Assume we have proven the assertion
for a certain $n \geq 2$. 
Let $(g_1,\ldots,g_n,g_{n+1})$ be some element in $\LG^{n+1}$. 

First, we generate the components $1$ to $n$.
For this, let $W$ be the ($n+1$)-restriction $R_{n+1}(V)$ of $V$.
By Lemma \ref{lem:restrict+rich}, $W$ is rich, 
hence -- by induction hypothesis -- we have 
$(g_1,\ldots,g_n) \in \pot\LG W q$ for some $q$. 
Again by Lemma \ref{lem:restrict+rich}, there is some $g\in \LG$
with $(g_1,\ldots,g_n, g) \in \pot\LG V q$.

Second, we generate the last component.
Since $\LG$ equals its own commutator subgroup,
we have some finite number $S$ and finitely many $g'_s, g''_s\in \LG$ with 
\bgl
\prod_{s=1}^S \: g'_s g''_s (g'_s)^{-1} (g''_s)^{-1} & = & g^{-1} g_{n+1}.
\egl 
Moreover, as shown above, there are certain $g'_{1,s}$ and $g''_{2,s}$ in $\LG$,
such that 
\bglklein
(g'_{1,s}, 1, 1, \ldots, 1, g'_s) \in \pot\LG V{q'} & \text{ and } &
(1, g''_{2,s}, 1, \ldots, 1, g''_s) \in \pot\LG V{q''}
\eglklein
for some $q', q''$.

Finally, we get 
\bgl
         (g_1,\ldots,g_n,g_{n+1}) 
 &  =  & (g_1,\ldots,g_n,g) \: (1,\ldots,1,g^{-1} g_{n+1}) \\
 &  =  & (g_1,\ldots,g_n,g) \cdot {} \\
 &     & \:\:\: \prod_{s=1}^S \Bigl(
                  (g'_{1,s}, 1, 1, \ldots, 1, g'_s) \: 
                  (1, g''_{2,s}, 1, \ldots, 1, g''_s) \cdot {} \\[-2.7ex]
 &     & \:\:\: \phantom{\prod_{s=1}^S \Bigl(}
                  (g'_{1,s}, 1, 1, \ldots, 1, g'_s)^{-1} \: 
                  (1, g''_{2,s}, 1, \ldots, 1, g''_s)^{-1} \Bigl)\\
 & \in & \pot\LG V{(q + 2(q'+q'')S)}.
\egl
Consequently, $\LG^{n+1} \teilmenge \bigcup_{q\in\N} \: \pot\LG V q$.
The opposite inclusion is trivial.
\eunum
If, additionally, $\LG$ has finite commutator length, 
the proof is 
analogous. Just observe that then  
we may choose $q,q',q'' \leq q(n)$ and $S \leq \cl(\LG)$ in the induction step. 
Setting $q(1) := 1$, $q(2) := 1$ and $q(n+1) := (1 + 4\:\cl(\LG)) \: q(n)$, we get
$\LG^{n+1} = \pot\LG V{q(n+1)}$.
\qed
\epf 
Directly from the proof we get a (very weak) upper bound for $q(n)$.
\bcorr
We can choose 
$q(n) \leq \bigl(1+4\:\cl(\LG)\bigr)^{n-2}$
for $n \geq 2$, provided $\LG$ has finite commutator length.
\ecorr
Moreover, we have 
\bcorr
\label{corr:allg_ss_full}
With the assumptions of Theorem \ref{thm:rich->full}, we have 
for every nonempty ordered subset $V$ of $\cxy n$ and for every 
$\LG$ having finite commutator length:
\bunum
\item
$\pot\LG V{q(n-\defi V)}$ is a subgroup of $\LG^n$ 
isomorphic to $\LG^{n-\defi V}$;
\item
$\pot\LG V{q(n-\defi V)}$ equals $\LG^n$ iff $V$ is rich or
$\LG$ is trivial;
\item
$\pot\LG V{q} = \pot\LG V{q(n-\defi V)}$ for all $q \geq q(n-\defi V)$.
\eunum
\ecorr
\bpf 
We may assume $V \neq \{(0,\ldots,0)\}$. Otherwise, the statements are obvious
setting $q(0) := 1$.

The basic idea is to extract from $V$ the independent and nontrivial 
components. For this, we divide the set of indices into equivalence classes,
whereas $i$ and $j$ are said to be equivalent iff for every element of $V$
its $i$- and $j$-component coincide.
Now, we define for every $v \in V$ a tuple $w$ 
just by dropping all the components that are $0$ for all $v \in V$
or that correspond to indices that are not minimal in their equivalence class. 
By construction, the set $W$ of all $w$ given this way is rich
and the number of components the elements of $W$ have, is $n - \defi V =: n'$.%
\footnote{On the one hand, $\defi V \leq n - n'$ by richness of $W$.
On the other hand, it is easy to see that every restriction $W'$ 
of $V$ having elements with more components than those in $W$, is not rich,
hence $\defi V \geq n - n'$. 
In fact, it has to contain still equivalent, but different components
or components that are always zero.}

By Theorem \ref{thm:rich->full}, we
have $\pot\LG W {q(n')} = \LG^{n'}$.
By construction, we see that 
$\pot\LG V{q(n')}$ consists precisely of all elements 
in $\LG^n$ that are constant on the equivalence classes above 
or that are the identity 
if they belong to the equivalence class built by the components 
being $0$ for all $v \in V$. Thus, $\pot\LG V{q(n')}$ is indeed
a subgroup of $\LG^n$.

By the group property of $\pot\LG V{q(n')}$, we have the
second inclusion relation in
$\pot\LG V{q(n')} \teilmenge \pot\LG V q = \pot\LG V{q(n')} \pot\LG V{(q-q(n'))}
\teilmenge \pot\LG V{q(n')}$ for all $q \geq q(n')$.
Finally, we see from the construction of $W$ that
$\pot\LG V{q(n')} \iso \pot\LG W{q(n')} = \LG^{n'}$.

The statement on the equality of 
$\pot\LG V{q(n-\defi V)}$ and $\LG^n$ is clear now.
\qed
\epf

\bcorr
The statements of Corollary \ref{corr:allg_ss_full}
are also true both in the category of Lie groups and that
of algebraic groups.
\ecorr
\bpf
By the construction in the corollary above, we see immediately that
$\pot\LG V{q(n')}$ is a submanifold of $\LG^n$ for Lie groups and is
closed in $\LG^n$ for algebraic groups.
\qed
\epf

\subsection{Abelian Groups}
Theorem \ref{thm:rich->full} is no longer true if we drop, e.g., the
``semisimplicity'' condition. For connected and compact $\LG$ it can be shown
that for 
\zgl{V := \{(1,1,0,0),\:(1,0,1,0),\:(0,1,0,1),\:(0,0,1,1)\}}\noindent
we have $\pot\LG V q = \LG^4$ iff $\LG$ is semisimple.
(The idea for the proof can be already found in \cite{diss,paper10}.)
This example (see Figure \ref{fig:stdex} on page \pageref{fig:stdex}) 
corresponds to the web introduced by Baez and Sawin \cite{d17}
that has been used widely to discuss problems arising in the theory
of webs.

However, at least for abelian groups we have a rather explicit description
of $\LG_V$:
\bprop
Let $\LG$ be some abelian group and let $n$ be some positive integer.
Then $\LG_V$ is a subgroup of $\LG^n$ for all 
nonempty ordered subsets $V$ of $\cxy n$
and we have 
\zgl{\LG_V = \LG \tensor \spann_\Z(V) = \LG^{\spann_\Z(V)}.}
This statement is also true in the category of real and complex Lie groups. 
\eprop
\bpf
This is a simple consequence of the fact that abelian groups are just 
$\Z$-modules.
\qed
\epf
Note, however, that the equality of $\LG_V$ and $\LG^n$ does {\em not}\/
only depend on $V$, but also on $\LG$. In fact, consider the example $n=4$ and
\zgl{V = \{(1,1,0,0),(1,0,1,0),(1,0,0,1),(0,1,1,0),(0,1,0,1),(0,0,1,1)\}.}\noindent
One checks immediately that 
\zgl{\spann_\Z(V) = 
      \{(z_1,z_2,z_3,z_4) \in \Z^4 \mid z_1 + z_2 + z_3 + z_4 \kong 0 \mod 2\}.}\noindent
Consequently, we have $(\Z_2)_V \echteteilmenge (\Z_2)^4$,
but $(\Z_3)_V = (\Z_3)^4$.
However, given a Lie group over $\R$ (or $\C$) the situation
is much nicer because of
\bcorr
For every positive integer $n$ and every ordered subset $V$ of $\cxy n$
we have 
\bgl
\R_V & = & \phantom(\spann_\R V \\
(U(1))_V & = & (\spann_\R V) / \Z^n.\\
\egl
In particular, we have $\LG_V = \LG^n$ iff $\spann_\R V = \R^n$ for 
$\LG = \R, U(1)$.
\ecorr
For the example 
$V = \{(1,1,0,0),\:(1,0,1,0),\:(0,1,0,1),\:(0,0,1,1)\}$ 
from the beginning of this subsection
and $\LG = \R$,
we have 
$\spann_\R V = \{(x_1,x_2,x_3,x_4) \mid x_1 - x_2 - x_3 + x_4 = 0\} 
  \teilmenge \R^4$.
In particular, $\dim \LG_V = 3$, but $n = 4$.

\subsection{More General Groups}
If we would restrict ourselves to the case of (connected) compact Lie groups,
we can always write $\LG$ as $(\LG_\he \kreuz U(1)^r)/\LN$, where
$\LG_\he$ is some semisimple compact Lie group, $r$ is some natural number and
$\LN$ is some discrete central subgroup of $\LG_\he \kreuz U(1)^r$. 
We know already the properties of $\LG_V$ for the semisimple and the
$U(1)$ case. Therefore it is natural to investigate, how
the subsets generated by $V$ in a product group are 
related to the corresponding subsets in the single groups
and for the factorized version as well. 
This question will be answered by the following propositions.
However, before we can state them, we have to introduce some 
(sloppy) notation. For $i=1,\ldots,k$ let $\LG_i$ be some group and 
$U_i$ be some subset of $(\LG_i)^n$ with a certain fixed $n$.
Then we can naturally identify 
\zgl{U_1 \kreuz \cdots \kreuz U_k \ident 
\bigl\{\bigl((g_{11},\ldots,g_{1n}),\:\ldots\:,(g_{k1},\ldots,g_{kn})\bigr)\bigr\}
\teilmenge (\LG_1)^n \kreuz \ldots \kreuz (\LG_k)^n}\noindent
with the subset
\zgl{\bigr\{\bigr((g_{11},\ldots,g_{k1}),\:\ldots\:,(g_{1n},\ldots,g_{kn})\bigr)\bigr\}}\noindent
of $(\LG_1 \kreuz \cdots \kreuz \LG_k)^n$.
Somewhat sloppily we denote this subset also by
$U_1 \kreuz \cdots \kreuz U_k$.
One sees immediately that properties being a subgroup, being a submanifold etc.\
are invariant under this identification -- its just an
isomorphism. Using this we have
\bprop
\label{prop:G_V(prod)_abstr}
Let $\LG_1, \ldots, \LG_k$ be some groups, $n$ be a positive integer and
$V$ be some ordered subset of $\cxy n$.

Then we have
$(\LG_1 \kreuz \cdots \kreuz \LG_k)_V 
  = (\LG_1)_V \kreuz \cdots \kreuz (\LG_k)_V$.
\eprop
\bpf
This is trivial, since the index $V$ just indicates
a certain product of elements without mixing any components.
\qed
\epf
\bprop
\label{prop:G_V(quot)_abstr}
Let $\LG$ be a group and $\LN$ be a normal subgroup of $\LG$.
Moreover, let $n$ be some positive integer and
$V$ be some ordered subset of $\cxy n$.

Then $(\LG/\LN)_V = \LG_V/\LN^n$, 
which is a subgroup of $(\LG/\LN)^n \iso \LG^n/\LN^n$.
\eprop
\bpf
Clear.
\qed
\epf
Altogether we have
\bthm
\label{thm:ulgru_abstr}
Let $\LG_\he$ be some group that equals its commutator subgroup and has 
finite commutator length.  
Let $\LG_\ab$ be some abelian group. Moreover, let $\LN$ be some
normal subgroup of $\LG_\he \kreuz \LG_\ab$.
Define $\LG := (\LG_\he \kreuz \LG_\ab)/\LN$.

Then there is a function $q:\N \nach \N_+$, such that 
$\pot\LG V{q(n-\defi V)}$ is a subgroup of $\LG^n$ 
for any nonempty ordered subset $V$ of $\cxy n$
and for every positive integer $n$. 
For rich $V$ we even have
\zgl{\pot\LG V{q(n)} = \bigl((\LG_\he)^n \kreuz (\LG_\ab)_V\bigr)/\LN^n.}
\ethm

\bpf
Choose $q(n)$ according 
to Theorem \ref{thm:rich->full} 
and Corollary \ref{corr:allg_ss_full}.
By the assertions above, we know that 
\zgl{\pot\LG V{q(n-\defi V)} 
       = \bigl(\pot{(\LG_\he)} V{q(n-\defi V)} \kreuz 
                       \pot{(\LG_\ab)} V{q(n-\defi V)}\bigr)/\LN^n}
which is a subgroup of $\LG^n$.
Additionally we used the group property
of $(\LG_\ab)_V$ to get $\pot{(\LG_\ab)} V {q(n)} = (\LG_\ab)_V$.
Moreover, the assertion for rich $V$ follows directly from
Theorem \ref{thm:rich->full}.
\qed
\epf
\bcorr
\label{corr:ulgru}
The statements of Proposition \ref{prop:G_V(prod)_abstr},
Proposition \ref{prop:G_V(quot)_abstr} and Theorem \ref{thm:ulgru_abstr}
(both for discrete $\LN$)
are true in the category of $\K$-Lie groups as well. 
Here, $\K$ is $\R$ or $\C$.
If, moreover, 
$\LG_\ab$ is connected,
then we have
\bgl
\codim_{\LG^n} \pot\LG V{q(n-\defi V)} 
 & = & \defi V \cdot \dim\LG_\he + 
             \phantom(\codim_{\R^n}\spann_\R V\phantom) \cdot \dim \LG_\ab\\
 & = & \defi V \cdot \dim\LG_\he + 
             (n - \dim\spann_\R V) \cdot \dim \LG_\ab
\egl
for all nonempty subsets $V$ of $\cxy n$.
\ecorr
\bpf
The transferability of the assertions above to the category of Lie groups
is clear. 
(Note that the discreteness of $\LN$ guarantees 
that the groups $\LG^n$ and $\LG^n/\LN^n$ are locally diffeomorphic.)
The codimension formula again is a consequence of the preceding statements.
Observe that every connected abelian Lie group over $\K$
is isomorphic to some $\R^r \kreuz U(1)^s$.
Here, of course, $r + s$ is even for $\K = \C$, such that 
$\R^r \kreuz U(1)^s$ can be regarded naturally as a complex Lie group.
\qed
\epf

\subsection{Application to Reductive Groups}
\bprop
\label{prop:ulgru}
For every positive integer $n$,
every nonempty subset $V$ of $\cxy n$ and every 
connected reductive linear algebraic group $\LG$ over $\R$ or $\C$,
the smallest subgroup generated by $\LG_V$ is a Lie subgroup of 
$\LG^n$. This smallest subgroup is given by $\pot\LG V q$ with $q$ 
being any integer $q\geq q(n-\defi V)$.
It equals $\LG^n$ iff 
$V$ is a generating system for $\R^n$, or $\LG$ is semisimple with rich $V$. 
\eprop
Note that every (real or complex) linear algebraic group 
has also a Lie group structure. In what follows, we will always
assume to use this Lie structure when speaking about linear algebraic groups.
Therefore, we may consider, e.g., 
Lie subgroups of linear algebraic groups, as we did in the proposition above. 
\bpf
\bunum
\item
Recall that every connected reductive linear algebraic group is 
the direct product of a connected 
semisimple Lie group and a connected abelian Lie group modulo
some discrete central subgroup. Additionally, every connected semisimple
Lie group coincides with its commutator subgroup. If it is even linear,
then it has finite center \cite{EMS41} and, therefore \cite{m2}, 
finite commutator length. 
Consequently, we are here precisely in
the setting of Theorem \ref{thm:ulgru_abstr} and Corollary \ref{corr:ulgru}. 
\item
Since the smallest subgroup of $\LG^n$ containing $\LG_V$
has to contain every $\pot\LG V q$ with $q \in \N$, 
it has to contain $\pot\LG V{q(n-\defi V)}$,
which, however, is already a Lie subgroup containing $\LG_V$. Hence, 
$\pot\LG V{q(n-\defi V)}$ is precisely that smallest subgroup generated 
by $\LG_V$. Corollary \ref{corr:ulgru} yields the assertion.
Since $\spann_\R V = \R^n$ implies that $V$ is rich, 
the condition for $\pot\LG V{q(n-\defi V)} = \LG^n$ is clear as well.
\qed
\eunum
\epf
Finally, we remark that every 
compact Lie group is a reductive linear algebraic group.

\section{Holonomical Independence}
\label{sect:holo_indep}
The aim of the present section is to investigate in detail
which parallel transports may be assigned by smooth connections 
to certain sets of paths. From now on, $\LG$ is a (real or complex) Lie group.

\bdf
\label{def:holo-iso}
Let $\gc = (\gamma_1,\ldots,\gamma_n)\teilmenge\Pf$ be some
tuple of paths and let $\triv$ be some ultralocal trivialization. 
Then $\gc$ is called
\bnum{3}
\item
\df{holonomically isolated} iff
\bunum
\item
the paths in $\gc$ are non-selfintersecting,
\item
$\bigl(\gc(0) \cup \gc(1)\bigr) \cap \inter \gc = \leeremenge$ and
\item
for every closed subset $K$ of $M$ with $K\cap\inter\gc = \leeremenge$,
for every $A\in\A$ and every
$(g_1,\ldots,g_n)\in\trv\A_\gc \teilmenge \LG^n$ 
there is an $A'\in\A$ such that 
\bunum
\item
$\trv h_{\gamma_i} (A') = g_i$ for $i = 1,\ldots,n$ and
\item
$A'$ and $A$ coincide on $K$;
\eunum
\eunum
\item
\df{holonomically independent} iff for every 
$(g_1,\ldots,g_n) \in \LG^n$ there is some $A\in\A$, such that
$\trv h_{\gamma_i}(A) = g_i$ for all $i=1,\ldots,n$;
\item
\df{strongly holonomically independent} iff it is
holonomically independent and holonomically isolated.
\enum
\edf
Note that $\inter \gc$ is defined by 
$\gc((0,1)) := \bigcup_i \gamma_i((0,1))$, 
which is generally not the interior of the image 
$\im \gc := \bigcup_i \gamma_i([0,1])$ of $\gc$.
Analogously, $\gc(0)$ and $\gc(1)$ are here the sets formed by the components
of the tuples $(\gamma_1(0),\ldots,\gamma_n(0))$
and $(\gamma_1(1),\ldots,\gamma_n(1))$, respectively.

We have obviously
\blem
The notions of Definition \ref{def:holo-iso} do not depend on the 
chosen trivialization $\triv$.
\elem

Before we come to the next statements, let us first discuss
the relevance of the preceding definition comparing, for
simplicity, the ``normal'' holonomy independence and the strong
holonomy independence.
In the case of the weaker, i.e.\ normal independence, it cannot 
a priori be excluded 
that products of independent sets of paths are not independent, even
if they are non-overlapping outside their endpoints:
Let $\vec g := (g_1,\ldots,g_n)\in\LG^n$ and let $\gc$ and $\gd$ be given.
Moreover, let $\vec g = \vec g' \vec g''$.
Now, it is, by independence, possible to find some $A'$ with
$h_\gc(A') = \vec g'$ and some $A''$ with $h_\gd(A'') = \vec g''$,
but this does not directly imply the existence of some 
$A'''$ with $h_{\gc\circ\gd}(A''') = \vec g' \vec g''$, maybe
just by setting $A''' \einschr{\im\gc} = A'$ and $A'''\einschr{\im\gd} = A''$.
To make this possible, i.e., to get a smooth $A'''$, 
one has to control at least the interface between
$\gc$ and $\gd$ what precisely is done for the strong form of 
holonomy independence. Here we explicitly demand that the independence
condition does not touch the values of the connections at the endpoints
of the paths. Making this heuristic discussion more precise will be 
the goal of the next few claims.

\blem
\label{lem:union(shi)=shi}
Let $\gc^{1}, \ldots, \gc^{J}$
be finitely many holonomically isolated tuples of paths 
and set%
\footnote{Here, the ``union'' of tuples is given in the natural way:
Simply list all components of all the $\gc^j$.
To be extremely precise: 
$\gc = (\gamma_1^1,\ldots,\gamma_{N_1}^1,\ldots\ldots,
    \gamma_1^J,\ldots,\gamma_{N_J}^J)$ 
if $\gc^j = (\gamma_1^j,\ldots,\gamma_{N_j}^j)$ for all $j$.}  
$\gc := \bigcup_j \gc^{j}$.
Assume, moreover, $\im\gc^{j} \cap \inter\gc^{j'}$ is
empty unless $j = j'$. Then we have:
\bunum
\item
$\gc$ is holonomically isolated with
$\trv\A_\gc = \grosskreuz_j \trv\A_{\gc^{j}}$ for every trivialization $\triv$. 
\item
$\gc$ is strongly holonomically independent, if each $\gc^{j}$ is.
\eunum
\elem
Note that if, additionally, all $\trv\A_{\gc^{j}}$ 
are independent of the trivialization $\triv$, then $\trv\A_\gc$ is so as well.
\bpf
\bunum
\item
$\gc$ is holonomically isolated.

The non-selfintersection property is obvious. 

Next, we have 
$(\gc^{j}(0) \cup \gc^{j}(1)) \cap \inter\gc^{j'} \teilmenge 
 \im\gc^{j} \cap \inter\gc^{j'} = \leeremenge$ for $j \neq j'$,
hence  
\bglklein
\bigl(\gc(0) \cup \gc(1)\bigr) \cap \inter \gc 
 & = & \bigcup_j \bigcup_{j'} \bigl(\gc^{j}(0) \cup \gc^{j}(1)\bigr) \cap \inter\gc^{j'} \\
 & = & \bigcup_j \bigl(\gc^{j}(0) \cup \gc^{j}(1)\bigr) \cap \inter\gc^{j} \\
 & = & \leeremenge
\eglklein
by the isolation property of every $\gc^{j}$.

The third condition can be proven inductively. 
For this, let there be given some closed $K \teilmenge M \setminus \inter\gc$,
some $A\in\A$ and some $g_i^j\in\LG$ corresponding to the paths
$\gamma_i^j \in \gc$, such that 
$\vec g \in \grosskreuz_j \A_{\gc^{j}}$.
(We fixed some trivialization $\triv$, but drop here
the corresponding superscripts for $\A_{\ldots}$ and $h_A$.)
We set $A_0 := A$ and choose some $A_j\in\A$ for $j = 1,\ldots,J$ such
that $h_{A_j}(\gamma_i^j) = g_i^j$ for all $i$ and 
that $A_j$ and $A_{j-1}$ coincide on $K \cup \bigcup_{j'\neq j} \im \gc^{j'}$. 
The last union is compact and, by assumption, disjoint to $\inter \gc^j$.
Now let $A' := A_J$. 
Since, of course, $\A_\gc \teilmenge \grosskreuz_j \A_{\gc^{j}}$,
we see that $A'$ has the desired properties for $\gc$ to be
holonomically isolated. 
But, moreover, the construction showed also that 
$\grosskreuz_j \A_{\gc^{j}} \teilmenge \A_\gc$, hence
their equality.
\item
The statement for the strong holonomy independence is now clear.
\qed
\eunum
\epf
\blem
\label{lem:shi-crit}
Let $A\in\A$ be some smooth connection and
$\gc = (\gamma_1,\ldots,\gamma_n)$ be some tuple of non-selfintersecting paths.
We assume that every $\gamma_i$ has an open neighbourhood 
$U_i \teilmenge M$ that contains $\gamma_i$ as an embedding.
Moreover, let all the $U_i$ be mutually disjoint.

Then $\gc$ is strongly holonomically independent.
\elem
\bpf
Let $K$ be closed with $K \cap \inter\gc = \leeremenge$.
Thus, $M\setminus K$ is an open neighbourhood for $\inter\gc$.
Hence, every
$U'_i := U_i \cap (M\setminus K)$ contains $\gamma_i\einschr{(0,1)}$ 
as an embedding. Moreover, $U'_i$ and $U'_j$ are disjoint for $i \neq j$.

Let now $A\in\A$ and choose some ultralocal trivialization $\triv$.
Then in the case of compact $\LG$ it is well known 
that for every $\vec g\in\LG^n$ 
there is some $A'\in\A$ with 
$\trv h_{A'}(\gc) = \vec g$,
such that $A'$ and $A$ coincide outside $\bigcup_i U'_i$.
The proof in the general case including non-compact $\LG$ 
is not really more difficult. For 
completeness it is given in Appendix \ref{app:hol_indep_non-comp}.
Moreover, by
$\bigcup_i U'_i = \bigcup_i (U_i \cap (M\setminus K)) \teilmenge M\setminus K$,
the two connections $A$ and $A'$ coincide
at least on $K$, whence $\gc$ is
holonomically isolated.
The holonomical independence is now obvious.
\qed
\epf

\blem
\label{lem:isol_teilweg->isol_weg}
Let $\gc$ be some finite tuple of non-selfintersecting paths in $M$, whereas
the image of $\gc$ is contained in some open set $U$,
such that $P$ restricted to $\pi^{-1}(U)$ is trivial.
Additionally, assume 
$\bigl(\gc(0)\cup\gc(1)\bigr) \cap \inter\gc = \leeremenge$.
Moreover, let $I = [I_-, I_+]$ be some closed interval in $[0,1]$ such that
$\gc([0,I_-])$, $\gc(\inter I)$ and $\gc([I_+,1])$ are mutually disjoint. 
Assume finally that $\gc\einschr I$ is holonomically isolated and
$\trv\A_\gc \teilmenge \trv\A_{\gc\einschr I}$ in some ultralocal trivialization
$\triv$ that is smooth on $U$.
Then we have:
\bnum{2}
\item
$\trv\A_\gc = \trv\A_{\gc\einschr I}$. 
\item
If $\trv\A_{\gc\einschr I}$ is a subgroup of $\LG^n$,
then $\gc$ is holonomically isolated.
\enum
\elem
\bpf
\bnum{2}
\item
$\trv\A_\gc = \trv\A_{\gc\einschr I}$

Since $\im\gc$ is contained in $U$ and there is some trivialization $\triv$
that is smooth on $U$, there is a connection
$A_0\in\A$ with trivial parallel transports w.r.t.\ $\triv$ along $\gc$.
Since $\gc\einschr I$ is holonomically isolated and since 
$\gc(\inter I) \ident \inter\gc\einschr I$ 
and the compact set $\gc([0,1]\setminus \inter I)$ are disjoint, we have 
for every $\vec g\in\trv\A_{\gc\einschr I}$ some connection
whose parallel transports 
on $\gc\einschr I$ equal $\vec g$ and
on $\gc\einschr{[0,I_-]}$ and
$\gc\einschr{[I_+,1]}$ equal those of $A_0$ being trivial.
Consequently, 
$\trv\A_{\gc\einschr I} \teilmenge \trv\A_\gc 
                        \teilmenge \trv\A_{\gc\einschr I}$.
\item
$\gc$ is holonomically isolated.

Before we start, we define 
$\gc_-$ by $\gc\einschr{[0,I_-]}$ and $\gc_+$ by $\gc\einschr{[I_+,1]}$.
\bunum
\item
$\trv h_A(\gc_-) \in \trv\A_{\gc\einschr I}$ for all $A\in\A$

Let $A$ be some smooth connection. Since, by assumption, the compact sets
$\im\gc_-$ and $\im\gc_+$ are disjoint, there is 
a smooth function $f$ being $1$ on $\im\gc_+$
with $\supp f \teilmenge U \setminus \im\gc_-$. 
Define $A'$ by $(1-f)A$ on $U$ 
(w.r.t.\ the fixed trivialization $\triv$ on $U$) 
and by $A$ outside. Of course, this is a connection
with $\trv h_{A'}(\gc_-) = \trv h_A(\gc_-)$ and $\trv h_{A'}(\gc_+) = (e_\LG,\ldots,e_\LG)$.
Consequently,
$\trv h_{A'}(\gc) = \trv h_{A'}(\gc_-) \: \trv h_{A'}(\gc\einschr I)$, hence
$\trv h_A(\gc_-) = \trv h_{A'}(\gc_-) = 
   \trv h_{A'}(\gc) \: \trv h_{A'}(\gc\einschr I)^{-1} 
   \in \trv\A_{\gc\einschr I}$
by the group property of $\trv\A_{\gc\einschr I}$ and
$\trv\A_\gc = \trv\A_{\gc\einschr I}$.
\item
$\trv h_A(\gc_+) \in \trv\A_{\gc\einschr I}$ for all $A\in\A$

This is shown completely analogously to 
$\trv h_A(\gc_-) \in \trv\A_{\gc\einschr I}$.
\item
$\trv h_A(\gc_-)^{-1} \: \trv\A_\gc \: \trv h_A(\gc_+)^{-1} 
       \teilmenge \trv\A_{\gc\einschr I}$ 
for all $A\in\A$

Using $\trv\A_\gc = \trv\A_{\gc\einschr I}$ and the group property of 
$\trv\A_{\gc\einschr I}$ again, we get the desired
relation.
\item
Isolation property of $\gc$

Let $K \teilmenge M$ be closed with 
$K \cap \inter \gc = \leeremenge$.
Define $K' := K \cup \im \gc_- \cup \im \gc_+$. 
Obviously, $K'$ is closed.
Let now $x\in K' \cap \inter\gc\einschr I$.
Since, by assumption, 
$K \cap \inter\gc = \leeremenge$, we have 
$x\in\gc([0,1]\setminus\inter I)$. This is a contradiction 
to $\gc(\inter I) \cap \gc([0,1]\setminus \inter I) = \leeremenge$. 
Hence, $K' \cap \inter\gc\einschr I = \leeremenge$.

Let now $A\in\A$ and $\vec g := (g_1,\ldots,g_n) \in \trv\A_\gc$.
Then, since --~as shown above~--
$\trv h_A(\gc_-)^{-1} \: \trv\A_\gc \: \trv h_A(\gc_+)^{-1} 
     \teilmenge \trv\A_{\gc\einschr I}$ 
and
since $\gc\einschr I$ is assumed holonomically isolated,
there is some $A''\in\A$, such that
\bunum
\item
$\trv h_{A''} (\gc\einschr I) = 
    \trv h_A (\gc\einschr{[0,I_-]})^{-1} \: 
    \vec g \:
    \trv h_A (\gc\einschr{[I_+,1]})^{-1}$ 
and
\item
$A''$ and $A$ coincide on $K'$.
\eunum
Consequently, $\trv h_\gc(A'') = \vec g$ as well as 
$A''$ and $A$ coincide on $K\teilmenge K'$.
\qed
\eunum
\enum
\epf

\section{Regularity and Consistent Parametrization}
\begin{figure}\begin{center}
\epsfig{figure=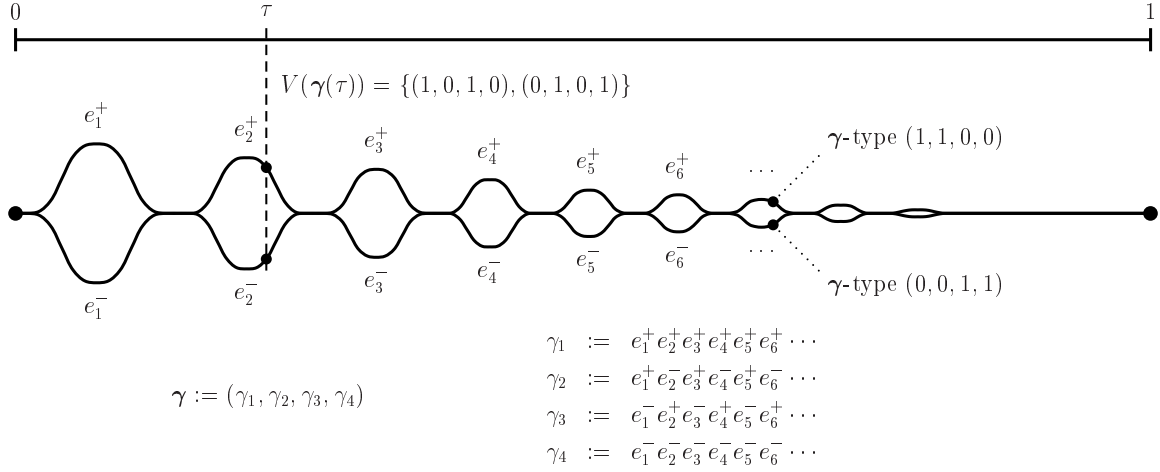,scale=0.77}
\caption{A special set of paths owing to Baez and Sawin \cite{d17}}
\label{fig:stdex}
\end{center}\end{figure}

In this section we get closer to the case of webs. First we recall 
(and slightly extend)
the notions of consistent parametrization and regularity \cite{d3}.
The latter one means, in particular, that, given a set of paths, 
around regular points 
there are no intersections between different paths,
unless they are locally identical.
The former one means that intersections of paths are allowed
only at those points where the paths under consideration have identical
parameter values. Afterwards we introduce
the type of points w.r.t.\ certain paths in generalization of
the similar notion settled by Baez and Sawin \cite{d3}.
\bdf
\label{def:consist}
An $n$-tuple $\gc=(\gamma_1,\ldots,\gamma_n)$ of paths
is called \df{consistently parametrized} iff 
we have for all $i,j = 1,\ldots,n$ 
\zgl{\gamma_i(t') = \gamma_j(t'') \breitrel\impliz t' = t''.} 
\edf
\bdf
\label{def:regular}
Let $\gc$ be some $n$-tuple of paths.
\bunum
\item
A point $x\in M$ is called \df{$\gc$-regular} iff $x$ is not the image
of an endpoint or nondifferentiable point of $\gc$ and there is a neighbourhood
of $x$ whose intersection with $\im \gc$ is an embedded interval.
\item
$\tau\in[0,1]$ is called 
$\gc$-regular iff $\gamma(\tau)$ is $\gc$-regular for all $\gamma\in\gc$.
\eunum
\edf
We have immediately
\blem
For every consistently parametrized $n$-tuple $\gc$ of paths in $M$,
the set of $\gc$-regular parameter values is open in $[0,1]$.
\elem
\bdf
\label{def:type}
Let $\gc = (\gamma_1,\ldots,\gamma_n)$ be some $n$-tuple of paths.
\bunum
\item
For every $x \in M$ we define 
the \df{$\gc$-type} $v(x)\in\cxy n$ of $x$ by
\zgl{v(x)_i := \begin{cases} 
                  1 & \text{if $x\in\im \gamma_i$} \\
                  0 & \text{if $x\nichtin\im \gamma_i$}
               \end{cases} .}
\item
For every consistently parametrized $\gc$ 
we define 
\zglklein{
V_\gc := \bigcup_{\text{$\tau\in[0,1]$, $\tau$ $\gc$-regular}} V(\gc(\tau)).
}
\eunum
\edf
For consistently parametrized $\gc$, obviously, $V(\gc(\tau))$ 
is the set of all $\gc$-types in $\gc(\tau)$.
Note, moreover, that in general the set $V_\gc$ of types in $\gc$ and the 
splitting $V(\gc)$ for $\gc$ do not coincide. For instance, we have
in the case of Figure \ref{fig:stdex}
\bgl
V_\gc & = & \{(1,1,0,0),(1,0,1,0),(0,1,0,1),(0,0,1,1)\}, \\ 
V(\gc) & = & \{(1,0,0,0),(0,1,0,0),(0,0,1,0),(0,0,0,1)\}.
\egl
\noindent
In a certain sense, $V_\gc$ is finer. $V(\gc)$ only looks whether two
whole paths are equal or not. $V_\gc$ looks closer at the image points 
of $\gc$.

Let us now study first consequences of the just introduced notions
for parallel transports. 

\blem
\label{lem:einbett_teil_PT}
Let $\gc$ be some consistently parametrized tuple of paths in $M$.
Assume that $\im\gc$ is contained in some open $U$, such that
$P(M,\LG)$ is trivial when restricted to the preimage of $U$.
Additionally, let $I \teilmenge [0,1]$ be some interval
whose endpoints are $\gc$-regular. 
Then we have $\trv\A_{\gc\einschr I} \teilmenge \trv\A_\gc$ for all
ultralocal trivializations $\triv$ that are smooth on $U$.
\elem
\bpf
Let $\vec g\in\trv\A_{\gc\einschr I}$, i.e.\
there is some smooth $A\in\A$, such that $\trv h_A(\gc\einschr I) = \vec g$.

Next, let $I = [I_-,I_+]$.
Since the set of $\gc$-regular parameter values
is open in $(0,1)$, there is some $\varepsilon>0$ 
such that $[I_- - 2 \varepsilon, I_-]$ and $[I_+, I_+ + 2 \varepsilon]$ 
are $\gc$-regular intervals. 
Set $J_{-\lambda} := [I_- - (\lambda+1)\varepsilon, I_- - \lambda\varepsilon]$
and $J_{+\lambda} := [I_+ + \lambda\varepsilon, I_+ + (\lambda+1)\varepsilon]$
for $\lambda \geq 0$. Finally, we set 
$J_{-\infty} := [0, I_- - 2\varepsilon]$ and
$J_{+\infty} := [I_+ + 2\varepsilon, 1]$.

Because $\gc$ is consistently parametrized and $\gc$ is continuous,
the sets $\gc(J_{-\infty})$, $\gc(I)$ and $\gc(J_{+\infty})$
are disjoint and compact. 
Consequently, there is some smooth function $f$ on $M$ 
that has support on $U \setminus \gc(I)$ and equals 
$1$ on $\gc(J_{-\infty}) \cup \gc(J_{+\infty})$.
We now define some connection $A'$ by $(1-f) A$ on $U$ and by $A$ outside
of $U$. Obviously, $A'\in\A$. 

Observe now, that, by regularity, 
two paths in $\gc$ coincide on $J_{\pm0}$ iff
they coincide on $J_{\pm1}$. Moreover, they coincide
iff their images have just a common point. 
Set 
$\gd = \{\delta_1,\ldots,\delta_l\} 
     := \gc\einschr{\inter J_{-1}} \cup \gc\einschr{\inter J_{+1}}$,
considered as sets.
Choose now some disjoint open sets $V_j$ in $U$ that are disjoint
to $\gc\bigl([0,1] \setminus (\inter J_{-1} \cup \inter J_{+1})\bigr)$,
such that there are certain closed intervals $I_j$ in $J_{-1}$ or $J_{+1}$,
respectively,
such that $V_j$ is a neighbourhood for $\delta_j(I_j)$
and that $\im \delta_j\einschr{I_j} \cap V_j$ is embedded into $V_j$. 
Then, by Proposition \ref{prop:holo_indep_nc}
there is some smooth connection $A''\in\A$, such that
\bunum
\item
$\trv h_{A''}(\gc\einschr{J_{-1}}) = \trv h_{A'}(\gc\einschr{J_{-0}})^{-1}$;
\item
$\trv h_{A''}(\gc\einschr{J_{+1}}) = \trv h_{A'}(\gc\einschr{J_{+0}})^{-1}$;
\item
$A'$ and $A''$ coincide outside $\bigcup V_j$.
\eunum
Altogether, we have 
\bgl
\trv h_{A''}(\gc) 
 & = & \trv h_{A''}(\gc\einschr{J_{-\infty}}) \:
       \trv h_{A''}(\gc\einschr{J_{-1}}) \:
       \trv h_{A''}(\gc\einschr{J_{-0}}) \:
       \trv h_{A''}(\gc\einschr{I}) \: 
\cdot \\ && \phantom{\trv h_{A''}(\gc\einschr{[0,I_-]}) \: \trv h_{A''}(\gc\einschr{J_{-1}}) \:} \cdot
       \trv h_{A''}(\gc\einschr{J_{+0}}) \:
       \trv h_{A''}(\gc\einschr{J_{+1}}) \:
       \trv h_{A''}(\gc\einschr{J_{+\infty}}) \\
 & = & \trv h_{A'}(\gc\einschr{J_{-\infty}}) \:
       \trv h_{A'}(\gc\einschr{J_{-0}})^{-1} \:
       \trv h_{A'}(\gc\einschr{J_{-0}}) \:
       \trv h_{A'}(\gc\einschr{I}) \: 
\cdot \\ && \phantom{\trv h_{A''}(\gc\einschr{[0,I_-]}) \: \trv h_{A''}(\gc\einschr{J_{-1}}) \:} \cdot
       \trv h_{A'}(\gc\einschr{J_{+0}}) \:
       \trv h_{A'}(\gc\einschr{J_{+0}})^{-1} \:
       \trv h_{A'}(\gc\einschr{J_{+\infty}}) \\
 & = & \trv h_{A}(\gc\einschr{I}) \\
 & = & \vec g.
\egl
Consequently, $\vec g \in \trv\A_\gc$.
\qed
\epf
\bprop
\label{prop:poss_values_for_conn}
Let $\gc:=(\gamma_1,\ldots,\gamma_n)$ 
be some consistently parametrized tuple of paths in $M$.
Assume that $\im\gc$ is contained in some open $U$, such that
$P(M,\LG)$ is trivial if restricted to the preimage of $U$.
Moreover, let $(\tau_k)$ be some finite, strictly increasing sequence 
of $\gc$-regular parameter values in $[0,1]$. 
Then we have, if $\triv$ is any ultralocal trivialization being smooth on $U$:
\bnum{2}
\item
If $\trv\A_\gc$ is a subgroup of $\LG^n$ 
and equals $\prod_{k} \LG_{V(\gc(\tau_k))}$, then $\gc$ is 
holonomically isolated.
\item
$\trv\A_\gc$ contains $\prod_{k} \LG_{V(\gc(\tau_k))}$.
\enum
\eprop
\bpf
Let us fix for all $k$ some $\tau_k^-$ and $\tau_k^+$ 
with $0 < \tau_k^- < \tau_k < \tau_k^+ < 1$, such that
every $\tau \in [\tau_k^-,\tau_k^+] =: I_k$ is $\gc$-regular.%
\footnote{Remember that $0$ and $1$ are not $\gc$-regular.}
Moreover we demand $\tau_{k-1}^+ < \tau_k^-$ for all $k\neq 1$.
Such $\tau_k^\pm$ always exist by the regularity of $\tau_k$.
Again by regularity we have $\gamma_i(\tau) = \gamma_j(\tau)$ with
$\tau \in I_k$ iff $\gamma_i(\tau_k) = \gamma_j(\tau_k)$, i.e., the
restrictions of two paths to $I_k$ are identical iff their images are
non-empty. This implies $V(\gc(\tau)) = V(\gc(\tau_k))$ (possibly
up to the ordering) for all $\tau \in I_k$,
hence $\LG_{V(\gc(\tau))} = \LG_{V(\gc(\tau_k))}$ 
by Lemma \ref{lem:split-gru-eig}.
Additionally, we define $I := \bigcup_k I_k$ and 
$J := \bigcup_{k\neq 1} [t_{k-1}^+,t_k^-]$, i.e., $J$ contains
all (closures of) intervals between the $I_k$s. Moreover, set
$J_0 := [0,t_1^-]$ and $J_1 := [t_{k_{\max}}^+,1]$. Finally, we
define the interval $L := I \cup J = [t_1^-,t_{k_{\max}}^+]$.

Let now $\gd^{k}$ be the 
set of all $\gamma_i\einschr{I_k}$, $i = 1,\ldots,n$,
and $\gd := \bigcup_k \gd^k$. Then $\im\delta$ is compact for every 
$\delta\in\gd$, and $(\im\delta') \cap (\im\delta'')$ is non-empty 
iff $\delta' = \delta''$ for $\delta',\delta'' \in \gd$.
Since every $\delta\in\gd$ is strongly holonomically independent by
Proposition \ref{prop:holo_indep_nc},
their union $\gd$ is so by Lemma \ref{lem:union(shi)=shi}. 

Now we are prepared for the proofs of the assertions in the proposition:
\bnum{3}
\item
Assume $\prod_{k} \LG_{V(\gc(\tau_k))} = \trv\A_\gc$ 
being a subgroup of $\LG^n$.

By consistent parametrization, the first two conditions for $\gc\einschr L$
to be holonomically isolated are fulfilled. To prove the third one,
let $K$ be some closed subset of $M$ with 
$K\cap\inter\gc\einschr L = \leeremenge$ and let $A\in\A$ be arbitrary. 
Since $\gc(J) \teilmenge \inter\gc\einschr L$ 
is compact, there is some smooth function $f$ on $M$
being $1$ on $\gc(J)$ with $\supp f \teilmenge U \cap (M\setminus K)$,
hence $f = 0$ on $K$.
We define $A_f$ to be the connection that coincides on 
$U$ with $(1-f) A$ and equals $A$ outside. 

Since $I_k$ is always $\gc$-regular,
Lemma \ref{lem:einbett_teil_PT} yields 
$\trv\A_{\gc\einschr L} \teilmenge \trv\A_{\gc} 
     = \prod_{k} \LG_{V(\gc(\tau_k))}$.
Then for every 
$(g_1,\ldots,g_n) \in \prod_k \LG_{V(\gc(\tau_k))}
    \obermenge \trv\A_{\gc\einschr L}$ 
we have 
certain $(g_{1,k},\ldots,g_{n,k})$ in $\LG_{V(\gc(\tau_k))}$ with
$(g_1,\ldots,g_n) = \prod_k (g_{1,k},\ldots,g_{n,k})$.
Since $\gd$ is strongly holonomically independent and 
$\inter\gd \teilmenge \bigcup _k \inter\gc\einschr{I_k}$, 
there is some $A' \in \A$ such that, in particular,
\bunum
\item
$\trv h_{\delta}(A') = g_{i,k}$ for all $i, k$ and $\delta\in\gd$ with 
$\delta = \gamma_i\einschr{I_k}$;
\item
$A'$ and $A_f$ coincide on $K \cup \gc(J)$. 
\eunum
$\trv h_{\delta}(A')$ is indeed well defined, because
$\gamma_{i'}\einschr{I_{k'}} = \delta = \gamma_{i}\einschr{I_{k}}$ 
implies $k' = k$, hence $\gamma_{i'}(\tau_{k}) = \gamma_{i}(\tau_{k})$
and thus $g_{i',k'} \ident g_{i',k} = g_{i,k}$ 
by the definition of $\LG_{V(\gc(\tau_k))}$.

Since $A_f$ is zero on $\gc(J)$,
the parallel transports along all subpaths of $\gamma_i$ for the
parameter intervals $[\tau_{k-1}^+,\tau_k^-]$ with $k\neq 1$ are $e_\LG$.
Hence, by construction,
\zglklein{\trv h_{\gamma_i\einschr L}(A') 
  = \prod_k \trv h_{\gamma_i\einschr{I_k}} (A') 
  = \prod_k g_{i,k} = g_i} for all $i$.
Consequently, we see first that 
$\trv\A_{\gc\einschr L} = \prod_k \LG_{V(\gc(\tau_k))} = \trv\A_\gc$,
and second that $\gc\einschr L$ is holonomically isolated.

Hence, $\gc$ is holonomically isolated by 
Lemma \ref{lem:isol_teilweg->isol_weg}.
\item
Now we show $\prod_k \LG_{V(\gc(\tau_k))} \teilmenge \trv\A_\gc$.

In contrast to the step above, we now choose some smooth function
$f$ on $M$ being $1$ on the full $\im\gc$ with $\supp f \teilmenge U$.
Analogously defining $A_f$ and choosing $A'$ we
get for every $(g_1,\ldots,g_n) \in \prod_k \LG_{V(\gc(\tau_k))}$
some $A'\in\A$ with $\trv h_{\gamma_i} (A') = g_i$, because
now $A'$ is trivial on $\gc(J_0 \cup J \cup J_1)$.
\qed
\enum
\epf

\section{Parallel Transports Along Webs}
We now recall the definition of tassels and webs 
owing to Baez and Sawin \cite{d3,d17}.
\bdf
\label{def:web}
\bunum
\item
A finite ordered set $T = \{c_1,\ldots,c_n\}$ 
of paths is called \df{tassel based on} 
$p\in\im T$ iff the following conditions are met:
\bnum{5}
\item 
$\im T$ lies in a contractible open subset of $M$.
\item 
$T$ can be consistently parametrized in such a way 
that $c_i(0) = p$ is the left endpoint of every path $c_i$.  
\item 
Two paths in $T$ that intersect at a point other than $p$
intersect at a point other than $p$ in every neighborhood of $p$.
\item 
For every neighbourhood $U$ of $p$, 
any $T$-type which occurs at some regular point in $\im T$ occurs at
some regular point in $U \cap \im T$.
\item 
No two paths in $T$ have the same image.
\enum
\item
A finite collection $\web = \web^1 \cup \dots \cup \web^k$ 
of tassels is called \df{web} iff for all $i \neq j$ the following
conditions are met:
\bnum{3}
\item 
Any path in the tassel $\web^i$ intersects any path in 
$\web^j$, if at all, only at their endpoints.
\item 
There is a neighborhood of each such intersection point 
whose intersection with $\im(\web^i \cup \web^j)$ 
is an embedded interval.
\item 
$\im \web^i$ does not contain the base of $\web^j$.
\enum
\eunum
\edf
Next, we list some important properties of webs that can 
be derived immediately from statements in \cite{d3}.
\bprop
\label{prop:eig(webs)}
For every web $w$ the set $[0,1]_\reg$ of $w$-regular parameter values 
is open and dense in $[0,1]$.
Moreover, 
the function $V(w(\cdot)) : [0,1]_\reg \nach \cxy{\elanz w}$, assigning
to every $w$-regular $\tau$ its splitting, is locally constant.
\eprop
\bpf
A slight modification of the proof of Lemma 1 in \cite{d3}
yields that for every $c_j\in w$ 
the set of all $\tau$ in $[0,1]$ with $w$-regular $c_j(\tau)$ 
is open and dense in $[0,1]$. Since the intersection of finitely many
open and dense subsets is again open and dense%
\footnote{Let $X_1, X_2$ be open and dense subsets in some 
topological space $X$.
Then, of course, $X_1 \cap X_2$ is open again. Assume $X_1 \cap X_2$ were
not dense in $X$. Then there is some $x\in X$ and some open
neighbourhood $U$ of $x$ in $X$, 
such that $U \cap (X_1 \cap X_2) = \leeremenge$.
Since $X_1$ is dense, there is some $x_1 \in U \cap X_1$. 
Since $X_2$ is dense, the open neighbourhood $U \cap X_1$ of $x_1$ 
must contain some $x_2 \in X_2$.
Consequently, 
$x_2 \in U \cap X_1 \cap X_2 = \leeremenge$.
Contradiction. The case of finitely many $X_i$ is now clear.}, 
$[0,1]_\reg$ is open and dense in $[0,1]$.
The second assertion is obvious.
\qed
\epf
\blem
\label{lem:tassel_rich}
For every web $w$ the set $V_w$ of $w$-types occurring in $w$ is rich.
\elem
\bpf
Since $w = \{c_1,\ldots,c_n\}$ is a web, for $i \neq j$ 
there is some $\tau\in[0,1]$ with $c_i(\tau) \neq c_j(\tau)$.
By Proposition \ref{prop:eig(webs)},
there is even a regular $\tau$ with this property,
thus $v(c_i(\tau)) \in V_w$.
By the consistent parametrization we have 
$v(c_i(\tau))_i = 1 \neq 0 = v(c_i(\tau))_j$,
i.e., the first richness condition is fulfilled.
The second is trivial.
\qed
\epf
For the following lemma, we still have to define the set
\bgl 
\was(w) &:=&
      \bigcap_{\tau\in(0,1]} \: \bigcup_{\sigma\in[0,\tau]_\reg} \bigl\{V(w(\sigma))\bigr\}
\egl\noindent
of all those (``regular'') 
splittings $V(w(\sigma))$ that appear in every neighbourhood
of $0$. Here, $I_\reg$ denotes the set of $w$-regular elements 
in an arbitrary interval $I\teilmenge[0,1]$.
\blem
\label{lem:types_in_V(w)}
Let $w$ be a web. 
Then for all $v\in V_w$ there is some $V\in\was(w)$ with $v\in V$.
In particular, $\was(w)$ is nonempty (if $w$ is nonempty).
\elem
\bpf
For every $v\in V_w$, by definition of a web, 
there is a sequence $\tau_i \gegen 0$ in $[0,1]_\reg$ 
with $v\in V(w(\tau_i)) \teilmenge \cxy{\elanz w}$ 
for all $i$. Since $\cxy{\elanz w}$ is finite, 
there is some $V \teilmenge \cxy{\elanz w}$
and some infinite subsequence $\tau_{i'} \gegen 0$ with
$V = V(w(\tau_{i'})) \ni v$ for all $i'$. By definition, $V\in\was(w)$.
\qed
\epf

\bcorr
\label{corr:rich(V_w)}
$\bigcup_{V\in\was(w)} V$ equals $V_w$ for every web $w$ and is rich.
\ecorr

Let us now state the main result of our article.
\bthm
\label{thm:mainthm}
Let $\LG$ be a connected reductive (real or complex) linear algebraic group
and let $T$ be some tassel.
Then for every $t\in (0,1]$ there is some $t'\in (0,t]$, such that
for every $0 \leq \tau \leq t'$ and every ultralocal trivialization $\triv$
\bunum
\item
$T\einschr{[\tau,t]}$ is holonomically isolated;
\item
$\trv\A_{T\einschr{[\tau,t]}}$ is a Lie subgroup of $\LG^n$ and equals
$\pot\LG{V}{q(\elanz T)}$. 
\eunum
\ethm
Recall that every linear algebraic group is a Lie group and that, 
in particular, every compact Lie group is a reductive linear algebraic
group. Hence, the assertion of the theorem above holds for all 
connected compact Lie groups $\LG$.

A part of the following proof is owing to \cite{d3}.
\bpf
\bunum
\item
Choice of $t'$

Denote the paths in $T$ by $c_1,\ldots,c_n$ and fix some 
ultralocal trivialization $\triv$ being smooth on some open neighbourhood
of $\im T$.
Fix, additionally, some ordering of $V := V_T$.
Since for every $v\in V$ there is some $V' \in \was(T)$ with $v \in V'$,
there is some finite sequence $V^{(s)}$ in $\was(T)$ with
$\prod_{s=1}^S \LG_{V^{(s)}} \obermenge \LG_{V}$.
Let now $(V^{(s)})_{s=1}^{Sq(n)} \teilmenge \cxy n$ be the $q(n)$-times 
repeated sequence of these elements in $\was(T)$
and set $t_{Sq(n)+1} := t$. 
Starting with $s = Sq(n)$, we choose inductively 
some $T$-regular $t_s \in (0,t_{s+1})$, such that $V(T(t_s)) = V^{(s)}$. 
By definition of $\was(T)$, such a $t_s$ always exist.
Finally, we choose some regular $t' < t_1$. 
\item
$\prod_{s=1}^{Sq(n)} \LG_{V^{(s)}} = \pot\LG V{q(n)}$

Let $v\in V^{(s)} \in \was(T)$ for some $s$. Then $v\in V$, i.e.\
$\LG_v \teilmenge \pot\LG V{q(n)}$. 
By the group property of the right-hand side, we have 
$\prod_{s=1}^{Sq(n)} \LG_{V^{(s)}} \teilmenge \pot\LG V{q(n)}$.
The opposite relation comes from 
$\prod_{s=1}^{S} \LG_{V^{(s)}} \obermenge \LG_V$ together with the 
definition of $(V^{(s)})$ as a $q(n)$-fold repetition of the first $S$
sets.
\item
$\trv\A_{T\einschr{[\tau,t]}} \obermenge \pot\LG{V}{q(n)}$ for 
$\tau \leq t'$

By the very definition of a tassel, $T\einschr{[\tau,t]}$ 
fulfills the requirements of Proposition \ref{prop:poss_values_for_conn}. 
Moreover, $(t_s)_{s=1}^{Sq(n)}$ is a strictly increasing sequence 
of $T$-regular values in $[\tau,t]$. 
Hence, by the construction of this sequence
and by the previous item, 
we have
$\trv\A_{T\einschr{[\tau,t]}} \obermenge \pot\LG{V}{q(n)}$.
\item
$\pot\LG{V}{q(n)}$ is a Lie subgroup of $\LG^n$

According to Lemma \ref{lem:tassel_rich},
the set of types in every tassel is rich. 
Proposition \ref{prop:ulgru} guarantees now that $\pot\LG V{q(n)}$ is even a 
Lie subgroup of $\LG^n$.
\item
$\trv\A_{T\einschr I} \teilmenge \pot\LG{V}{q(n)}$ for
every nontrivial interval $I \teilmenge [0,t]$,
where $T\einschr I$ consists of smooth immersive paths only

We have to prove that $\trv h_A(T\einschr I) \in \pot\LG V {q(n)}$ for all
$A\in\A$. For this, consider the map
\fktdefabgesetzt{\awpt}{I \kreuz I}{\LG^n.}%
      {(\tau_1,\tau_2)}%
      {\trv h_A(\tass\einschr{[\min I,\tau_1]})^{-1} \: 
        \trv h_A(\tass\einschr{[\min I,\tau_2]})}
Since $A$ and $\triv$ are smooth on $U$ containing $\im T$, 
the map $\awpt$ is smooth.
Moreover, if $(\tau_1,\tau_2)$ (or $(\tau_2,\tau_1)$) is some interval 
of $T$-regular parameters only 
(which implies that every element in it has the same type, say, $v$), 
then $H(\tau_1,\tau_2) \in \LG_v$, hence $\dot H(\tau_1,\tau_2) \in \LF\LG_v$ 
where the dot means differentiation w.r.t.\ the $\tau_2$-coordinate.
On the other hand, $\dot H(\tau_1,\tau_1)$ equals (up to the sign) 
$A\einschr{\tass(\tau_1)}(\dot\tass)$, where $\dot\tass$ is the 
$n$-tuple of tangential vectors on $\tass$ and 
$A\einschr{\tass(\tau_1)}$ is given naturally. 

Since we know that $\pot\LG V{q(n)}$ is a Lie group and since, as one sees
immediately, $\LG_v$ is Lie subgroup of $\pot\LG V{q(n)}$, the Lie algebra
$\LF\LG_v$ is a Lie subalgebra of $\LF(\pot\LG V{q(n)})$.
Consequently, $A\einschr{\tass(\cdot)}(\dot\tass)$
can be regarded as a smooth function
from $I$ to $\LF\LG^n$ with values in $\LF(\pot\LG V{q(n)})$ 
-- at least for those parameter values that are $T$-regular. 
However, since
the set of $T$-regular points is open and dense in $I$ 
and $A$ is smooth, $A\einschr{\tass(\cdot)}(\dot\tass)$ 
values in $\LF(\pot\LG V{q(n)})$ everywhere.
Since $\trv h_A(\tass\einschr I)$ is simply the path-ordered exponential of
$A\einschr{\tass(\cdot)}(\dot\tass)$ integrated along $I$, it is
contained in $\pot\LG V{q(n)}$.
\item
$\trv\A_{T\einschr{[\tau,t]}} \teilmenge \pot\LG{V}{q(n)}$ for 
$\tau \leq t'$

Since we consider piecewise smooth and immersive paths from the very beginning,
there are at most finitely many non-differentiability points in any
finite set of paths. Hence, we can decompose $[\tau,t]$ into 
a finite set $I_k$ of intervals, such that 
$T\einschr{[\tau,t]} = \prod T\einschr{I_k}$, where each
$T\einschr{I_k}$ consists of smooth and immersive paths only.  
Since $\trv h_A(T\einschr{[\tau,t]})$ is now the product of these
$\trv h_A(T\einschr{I_k})$, we get the assertion by the previous step
and the group property of $\pot\LG V {q(n)}$.
\item
$\tass\einschr{[\tau,t]}$ is holonomically isolated

This follows from Proposition \ref{prop:poss_values_for_conn},
because
\zglklein{\trv\A_{\tass\einschr{[\tau,t]}}
 = \pot\LG V{q(n)} = \prod_{s=1}^{Sq(n)} \LG_{V^{(s)}}} 
and $\pot\LG V{q(n)}$ is a subgroup of $\LG^n$.
\item
$\trv\A_{T\einschr{[\tau,t]}}$ is independent
of the ultralocal trivialization 

Let now $\triv'$ be an arbitrary ultralocal trivialization.
Then $\triv$ and $\triv'$ are related by a generalized gauge transform%
\footnote{A generalized gauge transform is a function from $M$ to $\LG$.}.
Consequently, we have 
\zgl{\trv[\triv']\A_{T\einschr{[\tau,t]}} = 
  (\vec g_{T(\tau)})^{-1} \: \trv\A_{T\einschr{[\tau,t]}} \: \vec g_{T(t)}}
for some $\vec g_{T(\tau)}, \vec g_{T(t)} \in \LG^n$. 
Since gauge transformations
depend only on the endpoints of paths, equal endpoints lead to equal
components in these two elements of $\LG^n$.
Hence, we have $\vec g_{T(\tau)} \in \LG_{V(T(\tau))}$ and 
$\vec g_{T(t)} \in \LG_{V(T(t))}$.

By Proposition \ref{prop:uhalbstet(split)} we get
$V(T(\tau)) \leq V(T(\tau'))$ and 
$V(T(t)) \leq V(T(t''))$ for some
$T$-regular $\tau', t'' \in [\tau,t]$.  
Thus, $\LG_{V(T(\tau))} \teilmenge \LG_{V(T(\tau'))}$
and $\LG_{V(T(t))} \teilmenge \LG_{V(T(t''))}$, whence
$\LG_{V(T(\tau))}$ and $\LG_{V(T(t))}$ are contained in $\LG_V$.
Since $\pot\LG V{q(n)}$ is a group, we have 
\zgl{
\trv[\triv']\A_{T\einschr{[\tau,t]}} 
 = (\vec g_{T(\tau)})^{-1} \: \trv\A_{T\einschr{[\tau,t]}} \: \vec g_{T(t)}
 = (\vec g_{T(\tau)})^{-1} \: \pot\LG V{q(n)} \: \vec g_{T(t)}
 = \pot\LG V{q(n)}.}
\qed
\eunum
\epf

\bcorr
Let $\LG$ be as in Theorem \ref{thm:mainthm}. Then for every web $w$
and every $t\in(0,1]$ there is some $t'\in(0,t]$, such that
for every $0 \leq \tau \leq t'$
\bunum
\item
$w\einschr{[\tau,t]}$ is holonomically isolated;
\item
$\A_{w\einschr{[\tau,t]}}$ is a Lie subgroup of $\LG^n$ 
with 
\zgl{
\A_{w\einschr{[\tau,t]}} = \grosskreuz_{i} \A_{T_i\einschr{[\tau,t]}} 
                         = \grosskreuz_{i} \pot\LG{V_{T_i}}{q(\elanz T_i)}
}
independent of the chosen ultralocal trivialization.
\eunum
Here, 
$w = T_1 \cup \ldots \cup T_W$ is a decomposition of $w$ into tassels $T_i$.
\ecorr
\bpf
This is an immediate consequence of 
the Theorem above and Lemma \ref{lem:union(shi)=shi}, since
two tassels share at most the endpoints of their paths with parameter
value $1$.
\qed
\epf
Setting $t$ to $1$ and $\tau$ to $0$, we get with the same notations
\bcorr
\label{corr:web_holoiso_Lie}
Let $\LG$ be as in Theorem \ref{thm:mainthm}. Then every web $w$
is holonomically isolated with
$\A_w = \grosskreuz_{i} \pot\LG{V_w}{q(\elanz w)}$ 
being a Lie subgroup of $\LG^{\elanz w}$. This again is independent of the
chosen trivialization.
\ecorr
\bpf
Every $v\in V_i := V_{T_i}$ 
can be interpreted as some $\dach v\in V_w$ simply by
adding zeros at all components that do not correspond to paths
in $T_i$. Since the only intersection points for tassels are 
at parameter value $1$ being not $w$-regular, each $\dach v \in V_w$, 
on the other hand, corresponds to precisely one $i$ and one $v\in V_i$. 
We have now 
\bgl
\LG_{\dach v} 
   & = & \{(e_\LG,\ldots,e_\LG)\} \kreuz \ldots \kreuz \{(e_\LG,\ldots,e_\LG)\} 
            \kreuz \LG_v \kreuz
\hspace*{\fill}\\&&\phantom{\{(e_\LG,\ldots,e_\LG)\}}\hspace*{\fill}{}\kreuz
     \{(e_\LG,\ldots,e_\LG)\} \kreuz \ldots \kreuz \{(e_\LG,\ldots,e_\LG)\}
\egl
with $\LG_v$ at component $i$.
Consequently, giving $V_w$ the ordering induced by the sequence of orderings
in $V_1,\ldots,V_n$, we have $\grosskreuz_{i} \LG_{V_i} = \LG_{V_w}$.
Moreover, since elements of $V_w$ lead to commuting $\LG_{\dach v}$s, if they 
correspond to different tassels, we have 
$\grosskreuz_{i} \pot\LG{V_i}q = \pot\LG{V_w} q$ for all $q\in\N$.
By $\elanz w \geq \elanz T_i$ for all $i$, we get
$q(\elanz w) \geq q(\elanz T_i)$ and thus
\zgl{\A_{w\einschr{[\tau,t]}} 
  = \grosskreuz_{i} \pot\LG{V_i}{q(\elanz T_i)}
  = \grosskreuz_{i} \pot\LG{V_i}{q(\elanz w)}
  = \pot\LG{V_w}{q(\elanz w)}} 
by the group property of each $\pot\LG{V_i}{q(\elanz T_i)}$.
\qed
\epf
Using Theorem \ref{thm:rich->full} we get
\bcorr
\label{corr:web_shi(ss)}
Let $\LG$ be as in Theorem \ref{thm:mainthm}.

If $\LG$ is semisimple, 
then every web is strongly holonomically independent.
\ecorr
These two corollaries yield, moreover, a new proof
for the denseness results of smooth connections as a subset
of generalized connections, differing from that presented
in \cite{paper10} and now including also a huge class of 
non-compact structure groups:
\bprop
\label{prop:A_dense}
Assume $\LG$ as in Theorem \ref{thm:mainthm} and let $\dim M \geq 2$.

Then, in the category of piecewise smooth and immersive paths, 
$\A$ is dense in $\Ab$ iff $\LG$ is semisimple.
\eprop
Here, $\Ab = \Hom(\Pf,\LG)$ is the set of all generalized connections
\cite{a30,diss}. Note that this definition is (for non-semisimple $\LG$)  
different from the definition by $\varprojlim_{w} \A_w$ in \cite{d3}
that uses $\A_w$ instead of $\Ab_w = \LG^{\elanz w}$.
\bpf
Recall \cite{paper10}
that $\A$ is dense in $\Ab$ if $\A_w = \LG^{\elanz w}$ for all webs $w$.
This is given for connected semisimple $\LG$ according to 
Corollary \ref{corr:web_shi(ss)}. On the other hand, recall that
$\A$ is not dense in $\Ab$ if $\A_w$ is not dense in $\LG^{\elanz w}$ 
for some web $w$. However, by Proposition \ref{prop:ulgru}, this is the case
if $\LG$ is not semisimple:
Observe first that there is always a web having type
$V$ with $\dim\spann_\R V < \elanz w$ 
(e.g., the web given in \cite{d17}; 
or see Figure \ref{fig:stdex} on page \pageref{fig:stdex}) 
and second that a Lie subgroup of non-zero codimension 
is never a dense subgroup.
\qed
\epf
Finally, we remark that Corollary \ref{corr:web_holoiso_Lie} above
completes a purely algebraic verification of the 
statement by Baez and Sawin in \cite{d3} that for connected and compact
groups the parallel transports along a tassel form a Lie subgroup
of $\LG^{\elanz T}$. Here, however, the assumptions have been
significantly weakened: $\LG$ can now be any connected
reductive linear algebraic group. This includes, in particular,
all classical semisimple Lie groups -- be they compact or not. Therefore,
our results may be extended also to, e.g., $\LG = Sl(2,\C)$ being relevant
for the non-compactified version of loop quantum gravity.
Already in the compact case, 
the assumption of the Lie subgroup property was crucial to initiate
a theory of well-defined generalized measures \cite{d3}.
Although a measure theory in the non-compact case is still more or less
on a speculative level, our results together with those in \cite{paper3} 
may give some hope that, after these problems
are solved for piecewise analytic paths, 
the measure theory can even be extended to the smooth category.

\section{Acknowledgements}
I am very grateful to Jerzy Lewandowski, Andrzej Oko\l\'ow and Matthias Schmidt 
for fruitful discussions. 
Moreover, I thank John Baez and Stephen Sawin for clarifying 
some point in their paper \cite{d3}.
Finally, I am very grateful that this work has been supported 
by the Reimar-L\"ust-Stipendium of the Max-Planck-Gesellschaft and 
in part by NSF grant PHY-0090091.

\anhangengl
\section{Holonomy Independence in the General Case}
\label{app:hol_indep_non-comp}
\bprop
\label{prop:holo_indep_nc}
Let $\LG$ be some connected Lie group and
let $\gc = \{\gamma_1,\ldots,\gamma_n\}$ 
be some finite set of non-selfintersecting paths in the manifold $M$.
Suppose that for every $\gamma_i\in\gc$ there is a closed interval 
$I_i\teilmenge [0,1]$ and some open neighbourhood 
$U_i$ of $\gamma_i(I_i)$ such that 
$\im\gamma_i \cap U_i \einbett U_i$ is an embedding. Assume
that all $U_i$ can be chosen mutually disjoint. 

Then for every ultralocal trivialization $\triv$,
for every $A\in\A$ and for all $g_1,\ldots,g_n \in \LG$ 
there is some $A'\in\A$ such that
\bunum
\item
$\trv h_{\gamma_i}(A') = g_i$ for all $i$;
\item
$A$ and $A'$ coincide outside $\bigcup_n U_n$.
\eunum
\eprop
\bpf
Observe first that by induction we may assume $n=1$. 
Moreover, if the statement of the proposition is true for one trivialization,
it is true for every trivialization. By the local triviality of the 
principal fibre bundle $P(M,G)$, we may assume that $P$ 
is trivial over $U := U_1$ and that $\gamma$ is smooth on $I := I_1$
(otherwise, shrink $U$ and $I$, if necessary). Hence,
we may fix some ultralocal trivialization $\triv$ being smooth on $U$.
In the following, however, we will simply write $h_A$ instead of $\trv h_A$.

We now proceed in two steps. First we modify $A$ such that 
it becomes ``zero'' in $\gamma(I)$ 
and second we modify it there to get the desired parallel transports.
\bunum
\item
First, since $M$ is a manifold, 
there is some smooth function $f$ on $M$ being $1$ on $\gamma(I)$ 
with $\supp f \teilmenge U$. 
We define $A_f$ to be equal $(1-f)A$ on $U$ w.r.t.\ $\triv$
and equal $A$ outside.
It is clear that $A_f$ is again a smooth connection.
To furnish the first part, we define 
$g' := h_{A_f}([0,\tau^-])^{-1} \: g \: h_{A_f}([\tau^+,1])^{-1}$.
\item
Recall that $g'$ --~as every element in a connected Lie group $\LG$~-- can be 
written as a product $g_1\cdots g_l$ of finitely many $g_j \in \exp(\Lieg)$,
where $\Lieg$ denotes the Lie algebra to $\LG$, i.e.,
$g_j = \exp(B_j)$ with $B_j \in \Lieg$.
Divide now $I$ into $2l-1$ intervals $I_1, I_{1\frac12}, I_2, \ldots, I_l$
by inserting $2l-2$ points  
and choose for all integer $j$ some
open neighbourhood $V_j \teilmenge U$ of some 
interior point in $I_j$, such that 
all these neighbourhoods are mutually disjoint and
$V_j \cap \im\gamma \teilmenge \gamma(I_j)$.
Choose finally (again for integer $j$) 
some smooth sections $f_j : M \nach T^\ast M$ 
with $\supp f_j \teilmenge V_j$ and fulfilling
$\int_{I_j} f_j(\gamma(t))_\mu \dot\gamma^\mu(t) \:\dd t = 1$.
Now we define $A'$ to be 
equal $A_f + \sum_j f_j B_j$ on $\bigcup_j V_j\teilmenge U$ and 
equal $A_f$ outside.
It is clear that $A'$ is again a smooth connection coinciding with $A$ 
outside $U$. Since $A'$ equals $f_j B_j$ on $V_j$ and since $B_j$ is a constant,
we have for integer $j$
\zglklein{h_{A'}(\gamma\einschr{I_j}) 
       = \exp\bigl(\int_{I_j} f_j(\gamma(t))_\mu B_j \dot\gamma^\mu(t) \:\dd t\bigr)
       = \exp(B_j) = g_j}
and thus
\bglklein
h_{A'}(\gamma) 
  & = & h_{A'}(\gamma\einschr{[0,\tau^-]}) \:
           \bigl(\prod_{\text{$j=1$, $2j\in\N$}}^{l} h_{A'}(\gamma\einschr{I_j})\bigr) \:
        h_{A'}(\gamma\einschr{[\tau^+,1]}) \\
  & = & h_{A_f}(\gamma\einschr{[0,\tau^-]}) \:
           \bigl(\prod_{\text{$j=1$}}^{l} g_j \bigr)  \:
        h_{A_f}(\gamma\einschr{[\tau^+,1]}) \\
  & = & h_{A_f}(\gamma\einschr{[0,\tau^-]}) \:
           g' \:
        h_{A_f}(\gamma\einschr{[\tau^+,1]}) \\
  & = & g.
\eglklein      
\qed
\eunum
\epf


\end{document}